\documentclass[useAMS,usenatbib]{mn2e}
\usepackage{graphicx}
\usepackage{latexsym}
\usepackage{psfig}
\usepackage{amssymb,amsmath}
\usepackage{subfigure}

\title[Mass segregation in star formation?]{Primordial mass segregation in simulations of star formation?}

\author[R. J. Parker et al.]{Richard  J. Parker$^{1}$\thanks{E-mail: R.J.Parker@ljmu.ac.uk}, James E. Dale$^{2}$ and Barbara Ercolano$^{2,3}$  \vspace*{0.1cm}\\
$^{1}$Astrophysics Research Institute, Liverpool John Moores University, 146 Brownlow Hill, Liverpool, L3 5RF, UK\\
$^{2}$Excellence Cluster `Universe', Boltzmannstra{\ss}e 2, 85748 Garching, Germany\\
$^{3}$Universit{\"a}ts-Sternwarte M{\"u}nchen, Scheinerstra{\ss}e 1, 81679 M{\"u}nchen, Germany}

\begin{document}

\pagerange{\pageref{firstpage}--\pageref{lastpage}} \pubyear{2014}

\maketitle

\label{firstpage}

\def\mnras{MNRAS}
\def\apj{ApJ}
\def\aj{AJ}
\def\aap{A\&A}
\def\apjl{ApJL}
\def\apjs{ApJS}
\def\araa{ARA\&A}
\def\pasj{PASJ}
 
\begin{abstract}
We take the end result of smoothed particle hydrodynamics (SPH) simulations of star formation which include feedback from photoionisation and stellar winds and evolve them for a further 10\,Myr using $N$-body simulations. We compare the evolution of each simulation to a control run without feedback, and to a run with photoionisation feedback only. In common with previous work, we find that the presence of feedback prevents the runaway growth of massive stars, and the resulting star-forming regions are less dense, and preserve their initial substructure for longer. The addition of stellar winds to the feedback produces only marginal differences compared to the simulations with just photoionisation feedback.

We search for mass segregation at different stages in the simulations; before feedback is switched on in the SPH runs, at the end of the SPH runs (before $N$-body integration) and during the $N$-body evolution. Whether a simulation is primordially mass segregated (i.e. before dynamical evolution) depends extensively on how mass segregation is defined, and different methods for measuring mass segregation give apparently contradictory results. Primordial mass segregation is also less common in the simulations when star formation occurs under the influence of feedback. Further dynamical mass segregation can also take place during the subsequent (gas-free) dynamical evolution. Taken together, our results suggest that extreme caution should be exercised when interpreting the spatial distribution of massive stars relative to low-mass stars in simulations.
\end{abstract}


\section{Introduction}

Understanding how and where stars form is one of the central pillars of astrophysics; star forming regions either form bound clusters \citep[][and references therein]{Lada03,Kruijssen12b} or (more usually) disperse into the Galactic disk and they are also the environment in which planetary systems are believed to form \citep[e.g.][]{Haisch01}. 

Hydrodynamical simulations \citep[e.g.][]{Bonnell08,Offner09,Girichidis11}, radiation-hydrodynamical simulations \citep[e.g.][]{Peters10,Bate12,Dale12a,Hansen12,Krumholz12} and radiation-magneto-hydrodynamic simulations \citep[e.g.][]{Myers14} of star formation make predictions for the outcome of star formation in dense, or clustered, environments. Ideally, we would like to compare the outcome of simulations of star formation to observations of young star-forming regions to search for similarities in stellar mass functions \citep{Bonnell97,Bate09,Krumholz12}, binary and multiplicity properties \citep{Donate04b,Goodwin04b,Offner10,Bate12} and in the spatial distributions of stars \citep{Schmeja06,Girichidis12}, including mass segregation \citep{Moeckel09b,Maschberger11,Kirk14,Myers14}.

One drawback of hydrodynamical simulations is that they do not follow the full dynamical evolution of a star forming region, either until it forms a cluster, or disperses altogether. This can be remedied slightly by evolving the simulation using a pure $N$-body code \citep[e.g.][]{Moeckel10,Moeckel12,Parker13a}. Whilst this approach cannot accurately model the gas left over from star formation, recent studies \citep[e.g.][]{Offner09,Smith11,Kruijssen12a} suggest that the removal of gas does not strongly affect the subsequent evolution of the cluster, due to high local star formation efficiencies. For this reason, the classical picture of a cluster becoming unbound following gas removal \citep{Tutukov78,Lada84,Goodwin06} may not be valid.

In a previous paper \citep{Parker13a} we took the outcome of five pairs of smoothed particle hydrodynamics (SPH) simulations and evolved them forward in time using an $N$-body integrator. In each pair, one simulation formed stars under the influence of photoionisation feedback, and the other was a control run without feedback. The differences in the SPH calculation were largely limited to differences in the mass functions; the run without feedback formed fewer stars, but they had higher masses and higher stellar densities. The runs with feedback were less dense, and as a result the clusters that formed retained structure for longer, due to their longer relaxation times \citep{Parker13a}.

In this paper, we take recent simulations by \citet{Dale14} which include a further source of feedback -- namely stellar winds as well as photoionisation feedback -- and follow their dynamical evolution for a further 10\,Myr using $N$-body simulations. As in \citet{Parker13a} we determine the evolution of their spatial distributions, local stellar density and fraction of bound stars. We also look for mass segregation, both before and during the subsequent $N$-body evolution. Recently, \citet{Kirk14} and \citet{Myers14} have uncovered evidence for primordial mass segregation in their hydrodynamical and radiation-magneto-hydrodynamic simulations, respectively -- see also \citet{Maschberger11} who find a similar result in an analysis of the \citet{Bonnell08} simulation of star formation. 

Observationally, mass segregation has been found in some young star clusters \citep[e.g.\,\,the Orion Nebula Cluster --][]{Hillenbrand98,Allison09a}, but not in other regions, including those that contain massive ($>$8\,M$_\odot$) stars \citep{Wright14} and those that do not \citep{Kirk10,Parker11b,Parker12c}. This begs the question of whether mass segregation is likely to be a primordial outcome of star formation \citep{Bonnell98,Kirk14,Myers14}, whether it is dynamical \citep{Allison09b,Allison10,Parker14b}, or some combination of the two \citep{Moeckel09b}. If none, or very little, mass segregation occurs in simulations of massive star formation, then the most likely scenario is likely to be that it is a predominantly dynamical process.

The paper is organised as follows. In Section~\ref{method} we provide a description of the SPH simulations from \citet{Dale12a,Dale13,Dale14}, and the set-up of the subsequent $N$-body simulations. We describe our results in Section~\ref{results}, we provide a discussion in Section~\ref{discuss}, and we conclude in Section~\ref{conclude}.

\section{Initial conditions}
\label{method}

In \citet{Parker13a}, we used as our starting conditions the results of SPH simulations of star formation in a parameter space of molecular clouds presented in \citet{Dale12a} and \citet{Dale13}. In these simulations, the influence of photoionising radiation from O--type stars was included according to the algorithm presented in \citet*{Dale07} and \citet{Dale12a}. \citet{Dale12a,Dale13} also present a control run of each simulation in which feedback was switched off. The SPH study has since been extended to include momentum-driven stellar winds as described in \citet{Dale08}, in addition to photoionisation. The simulations are terminated after feedback from the O--type stars has been active for 3\,Myr, since this is the time when these stars will begin to expire as supernovae. 

The principal results of the new SPH study are described in \citet{Dale14}. In general, the additional influence of winds on top of photoionisation is modest, except at very early times, when the winds aid the expanding HII regions in clearing dense gas from the deep potential wells in which the O--stars are situated. Nevertheless, there are differences in the numbers and distributions of stars formed in the dual-feedback simulations when compared to the ionisation-only runs.

In Table~\ref{cluster_props} we summarise the results from the SPH studies, including the simulations with stellar winds from the new study by \citet{Dale14}. We list each simulation triplet; the run without feedback first (`a'), the run with photoionisation feedback only (`b') and the run with photoionisation and stellar winds (`c'). For each triplet, we list the initial cloud virial ratio, $\alpha_{\rm vir}$, cloud mass, $M_{\rm cloud}$, the number of stars formed at the end of each simulation, $N_{\rm stars}$, the final stellar mass in the simulation, $M_{\rm {\bf region}}$, and the spatial structure as measured by the $\mathcal{Q}$--parameter \citep{Cartwright04} at the point feedback was switched on, and at the end of the SPH simulation.

\begin{table*}
\caption[bf]{A summary of the five different triplets of smoothed particle hydrodynamics (SPH) simulations used as the input initial conditions of our $N$-body integrations. The values in the columns are: the simulation number, the corresponding Run ID from \citet[][D12]{Dale12a}, \citet[][D13]{Dale13} or \citet[][D14]{Dale14}, the type of feedback in the SPH simulation (none, photoionisation only, or photoionisation and stellar winds), the paper reference,  the initial virial ratio of the original clouds $\alpha_{\rm init}^{\rm SPH}$ (to distinguish bound from unbound clouds), the initial radius of the cloud in the SPH simulation ($R_{\rm cloud}$), the initial mass of the cloud ($M_{\rm cloud}$),   the number of stars that have formed at the end of the SPH simulation ($N_{\rm stars}$), the mass of this star-forming region ($M_{\rm region}$), the $\mathcal{Q}$-parameter in the SPH simulation at the time feedback is initiated in the feedback runs ($\mathcal{Q}_{\rm init}^{\rm SPH}$), and the final $\mathcal{Q}$-parameter in the SPH simulations ($\mathcal{Q}_{\rm fin}^{\rm SPH}$).}
\begin{center}
\begin{tabular}{|c|c|c|c|c|c|c|c|c|c|c|}
\hline 
Sim. No. & Run ID & Feedback & Ref.  & $\alpha_{\rm init}^{\rm SPH}$ &$R_{\rm cloud}$ & $M_{\rm cloud}$ & $N_{\rm stars}$ & $M_{\rm region}$ & $\mathcal{Q}_{\rm init}^{\rm SPH}$ & $\mathcal{Q}_{\rm fin}^{\rm SPH}$\\
\hline
1(a) & J & None & D12 & 0.7 & 5\,pc & 10 000\,M$_\odot$ & 578 & 3207\,M$_\odot$ &0.53&0.49\\
1(b) & J & Photoionisation & D12  & 0.7 & 5\,pc & 10 000\,M$_\odot$ & 685 & 2205\,M$_\odot$ &0.53&0.60\\
1(c) & J & Photoionisation + wind & D14 & 0.7 & 5\,pc & 10 000\,M$_\odot$ & 564 & 2186\,M$_\odot$ &0.53&0.70 \\
\hline
2(a) & I & None & D12 & 0.7 & 10\,pc & 10 000\,M$_\odot$ & 186 & 1270\,M$_\odot$ &0.42&0.72\\
2(b) & I & Photoionisation & D12 & 0.7 & 10\,pc & 10 000\,M$_\odot$ & 168 & 805\,M$_\odot$ &0.42&0.38\\
2(c) & I & Photoionisation + wind & D14 & 0.7 & 10\,pc & 10 000\,M$_\odot$ & 132 & 766\,M$_\odot$ &0.42& 0.49\\
\hline 
3(a) & UF & None & D13 & 2.3 & 10\,pc & 30 000\,M$_\odot$ & 66 & 1392\,M$_\odot$ &0.59&0.77\\
3(b) & UF & Photoionisation & D13 & 2.3 & 10\,pc & 30 000\,M$_\odot$ & 76 & 836\,M$_\odot$ &0.59&0.55\\
3(c) & UF & Photoionisation + wind & D14 & 2.3 & 10\,pc & 30 000\,M$_\odot$ & 93 & 841\,M$_\odot$ &0.59& 0.49\\
\hline
4(a) & UP & None & D13 & 2.3 & 2.5\,pc & 10 000\,M$_\odot$ & 340 & 2718\,M$_\odot$ &0.47&0.49\\
4(b) & UP & Photoionisation & D13 & 2.3 & 2.5\,pc & 10 000\,M$_\odot$ & 346 & 1957\,M$_\odot$ &0.47&0.57\\
4(c) & UP & Photoionisation + wind & D14 & 2.3 & 2.5\,pc & 10 000\,M$_\odot$ & 343 & 1926\,M$_\odot$ &0.47& 0.64\\
\hline
5(a) & UQ & None & D13 & 2.3 & 5\,pc & 10 000\,M$_\odot$ & 48 & 723\,M$_\odot$ &0.42&0.70\\
5(b) & UQ & Photoionisation & D13 & 2.3 & 5\,pc & 10 000\,M$_\odot$ & 80 & 648\,M$_\odot$ &0.42&0.46\\
5(c) & UQ & Photoionisation + wind & D14 & 2.3 & 5\,pc & 10 000\,M$_\odot$ & 77 & 594\,M$_\odot$ &0.42& 0.45\\
\hline
\end{tabular}
\end{center}
\label{cluster_props}
\end{table*}

\subsection{$N$-body evolution}

We take the final states of five triplets of simulations from  \citet{Dale12a}, \citet{Dale13} and \citet{Dale14} and assume that the combination of the first supernova and stellar winds instantaneously removes any remaining gas from both the feedback and non-feedback simulations, and evolve the resulting gas-free systems with an $N$-body code. 

We evolve the clusters using the  $4^{\rm th}$ order Hermite-scheme integrator \texttt{kira}  within the Starlab environment \citep[e.g.][]{Zwart99,Zwart01}. We take the masses, positions and velocities of the sink-particles from the SPH simulations and place these directly into the $N$-body integrator. In the majority of the SPH runs the stars are in virial equilibrium, or slightly sub-virial, at the end of the simulation (i.e. the initial conditions for the $N$-body integration). However, the initial conditions for simulations UF, UP and UQ were globally unbound, so that one might expect the stars and clusters formed to be in an unbound configuration. In fact, some parts of the globally unbound clouds become bound due to high velocity gas flows colliding and radiating away kinetic energy, which tends to occur in the dense areas of the clouds where most of the stars form. Therefore, in practice, these unbound clouds form stars that are roughly in virial equilibrium (with virial ratios ranging from 0.4 -- 0.7), apart from Run UF (with feedback), which is highly supervirial, with a virial ratio of 1.9. 

The simulation triplets are then evolved for 10\,Myr, without a background gas potential. The simulations contain several stars with masses $>~20$\,M$_\odot$ which are likely to evolve over the 10\,Myr duration of the $N$-body integration. For this reason, we use the \texttt{SeBa} stellar evolution package in the Starlab environment \citep{Zwart96,Zwart12}, which provides look-up tables for the evolution of stars according to the time dependent mass-radius relations in \citet*{Eggleton89} and \citet{Tout96}. Typically, \texttt{SeBa} updates the evolutionary status of stars on shorter timescales than the timestep in the \texttt{kira} integrator, although for extremely close systems a lag of up to one timestep can occur.  

\section{Results}
\label{results}

\begin{table*}
\caption[bf]{A summary of the results of our $N$-body integrations. The values in the columns are: the simulation number and corresponding Run ID from \citet{Dale12a}, \citet{Dale13} or \citet{Dale14} (`a' corresponds to the SPH run with no feedback switched on,  `b' corresponds to the SPH run with photoionisation feedback and `c' corresponds to the run with both photoionisation and stellar wind feedback), the initial mass of this star-forming region before $N$-body integration ($M_{\rm region, i}$), the final mass after 10\,Myr of $N$-body integration ($M_{\rm region, f}$),  the initial and final $\mathcal{Q}$-parameters ($\mathcal{Q}_{\rm i}$ and $\mathcal{Q}_{\rm f}$), initial and final median surface densities ($\Sigma_{\rm i}$ and $\Sigma_{\rm f}$) and the initial and final fractions of bound stars ($f_{\rm bound,i}$ and $f_{\rm bound,f}$).}
\begin{center}
\hspace*{-0.75cm}\begin{tabular}{|c|c|c|c|c|c|c|c|c|}
\hline 
Sim. No. & $M_{\rm region,  i}$ & $M_{\rm region, f}$ & $\mathcal{Q}_{\rm i}$ & $\mathcal{Q}_{\rm f}$ &  $\Sigma_{\rm i}$ & $\Sigma_{\rm f}$ & $f_{\rm bound,i}$ & $f_{\rm bound,f}$ \\
\hline
J, 1(a) & 3207\,M$_\odot$ & 2531\,M$_\odot$  & 0.49 & 1.91 & 4518\,stars\,pc$^{-2}$ & 0.4\,stars\,pc$^{-2}$  & 0.96 & 0.57 \\
J, 1(b) & 2205\,M$_\odot$ & 1857\,M$_\odot$  & 0.60 & 1.89 & 141\,stars\,pc$^{-2}$ & 2\,stars\,pc$^{-2}$  & 0.90 & 0.77 \\
J, 1(c) & 2186\,M$_\odot$ & 1879\,M$_\odot$  &  0.70 & 1.50  & 51\,stars\,pc$^{-2}$ & 2.5\,stars\,pc$^{-2}$  & 0.84 & 0.71\\
\hline
I, 2(a) & 1271\,M$_\odot$ & 751\,M$_\odot$  & 0.72 & 1.39 & 102\,stars\,pc$^{-2}$ & 0.3\,stars\,pc$^{-2}$  & 0.78 & 0.50 \\
I, 2(b) & 805\,M$_\odot$ & 640\,M$_\odot$  & 0.38 & 0.79 & 83\,stars\,pc$^{-2}$ & 0.1\,stars\,pc$^{-2}$  & 0.74 & 0.34 \\
I, 2(c) & 766\,M$_\odot$ & 591\,M$_\odot$  & 0.49  & 0.60  & 7.2\,stars\,pc$^{-2}$ & 0.2\,stars\,pc$^{-2}$  & 0.50  & 0.33\\
\hline
UF, 3(a) & 1392\,M$_\odot$ & 410\,M$_\odot$  & 0.77 & 1.01 & 6\,stars\,pc$^{-2}$ & 0.01\,stars\,pc$^{-2}$  & 0.76 & 0.25 \\
UF, 3(b) & 836\,M$_\odot$ & 511\,M$_\odot$  & 0.55 & 0.74 & 0.6\,stars\,pc$^{-2}$ & 0.01\,stars\,pc$^{-2}$  & 0.46 & 0.22 \\
UF, 3(c) & 841\,M$_\odot$ & 608\,M$_\odot$  & 0.49 & 0.73 & 0.5\,stars\,pc$^{-2}$ & 0.02\,stars\,pc$^{-2}$  & 0.44  & 0.15 \\
\hline
UP, 4(a) & 2718\,M$_\odot$ & 1765\,M$_\odot$  & 0.49 & 1.40 & 250\,stars\,pc$^{-2}$ & 0.2\,stars\,pc$^{-2}$  & 0.84  & 0.41 \\
UP, 4(b) & 1957\,M$_\odot$ & 1587\,M$_\odot$  & 0.57 & 1.27 & 24\,stars\,pc$^{-2}$ & 0.2\,stars\,pc$^{-2}$  & 0.73 & 0.44 \\
UP, 4(c) & 1926\,M$_\odot$ & 1569\,M$_\odot$  & 0.64 & 1.01 & 16\,stars\,pc$^{-2}$ & 0.2\,stars\,pc$^{-2}$  & 0.73 & 0.27 \\
\hline
UQ, 5(a) & 723\,M$_\odot$ & 337\,M$_\odot$  & 0.70 & 0.93 & 6\,stars\,pc$^{-2}$ & 0.02\,stars\,pc$^{-2}$  &  0.71 & 0.31 \\
UQ, 5(b) & 648\,M$_\odot$ & 485\,M$_\odot$  & 0.46 & 0.72 & 2\,stars\,pc$^{-2}$ & 0.04\,stars\,pc$^{-2}$  & 0.56 & 0.30 \\
UQ, 5(c) & 594\,M$_\odot$ & 408\,M$_\odot$  & 0.45 & 0.68 & 4\,stars\,pc$^{-2}$ & 0.02\,stars\,pc$^{-2}$  & 0.51 & 0.17 \\
\hline
\end{tabular}
\end{center}
\label{cluster_evol}
\end{table*}

\begin{figure}
\begin{center}
\rotatebox{270}{\includegraphics[scale=0.4]{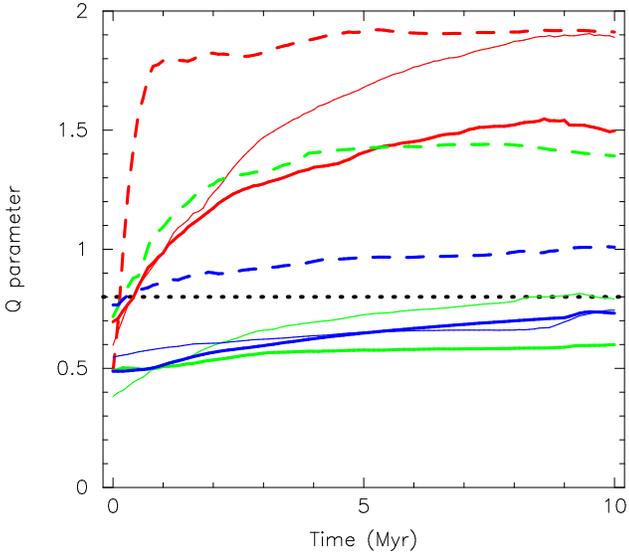}}
\end{center}
\caption[bf]{Evolution of the $\mathcal{Q}$-parameter for three of the five triplets of simulations (we omit runs UP and UQ from the plot for clarity, but their behaviour follows the same pattern). The boundary between a centrally concentrated, radially smooth distribution ($\mathcal{Q} > 0.8$) and a substructured distribution ($\mathcal{Q} < 0.8$) is shown by the dotted line. 
Simulations \emph{with} feedback are shown by the solid lines; those with ionization feedback only \citep[presented in][]{Parker13a} are shown by the thinner lines, and those with ionization feedback and stellar winds are shown by the thicker lines. Simulations with no feedback are shown by the dashed lines. The red lines are Runs J, green lines are Runs I, dark blue lines are Runs UF. Runs with feedback preserve structure for longer as the star-forming region evolves, and two simulations with feedback (I and UF) remain substructured throughout.}
\label{Qpar_all}
\end{figure}

In this Section we describe the $N$-body evolution of spatial structure, local surface density and the fraction of bound stars, before using three different methods to search for mass segregation. A summary of the evolution of structure, density and bound fraction, as well as the mass loss due to stellar evolution, is presented in Table~\ref{cluster_evol}.

\citet{Dale12a,Dale13} noted that the absence of feedback in the SPH calculations results in a more top-heavy IMF, as the most massive stars do not have their growth regulated by feedback. Photoionisation feedback reduces the mass of the star-forming region, and increases the number of stars formed. In general, the addition of stellar winds slightly reduces the number of stars compared to photoionisation feedback alone, and also slightly reduces the mass of the region (although there are exceptions, such as Run~UF). In all cases, the photoionisation+wind models reduce the initial stellar surface density with respect to the runs with photoionisation feedback only, and those with no feedback at all.

\subsection{Evolution of spatial structure}

We examine the evolution of the structure of the simulated regions over the duration of the $N$-body integration, using the $\mathcal{Q}$-parameter. The $\mathcal{Q}$-parameter was pioneered by \citet{Cartwright04,Cartwright09b} and combines the normalised mean edge length of the minimum spanning tree of all the stars in the 
region, $\bar{m}$, with the normalised correlation length between all stars in the region, 
$\bar{s}$. The level of substructure is determined by the following equation:
\begin{equation}
\mathcal{Q} = \frac{\bar{m}}{\bar{s}}.
\end{equation}
A substructured association or region has $\mathcal{Q}<0.8$, whereas a smooth, centrally concentrated cluster has $\mathcal{Q}>0.8$. The $\mathcal{Q}$-parameter has the advantage of being independent of the density of the star forming region, and purely measures 
the level of substructure present. The original formulation of the $\mathcal{Q}$-parameter assumes the region is spherical, but can be altered to take into account the effects of elongation \citep{Bastian09,Cartwright09a}. 
 
In Fig.~\ref{Qpar_all} we compare the evolution of the $\mathcal{Q}$-parameter with time in three of the five triplets of simulations (we omit Runs UP and UQ from the plot for clarity). The simulations that formed with feedback are shown by the solid lines; the simulations with ionisation feedback only that were presented in \citet{Parker13a} are shown by the thinner lines, and the simulations with both ionisation and stellar winds feedback are shown by the thicker lines. The simulations that formed without feedback are shown by the dashed lines. The colours correspond to the following simulations; red--Run~J, green--Run~I, dark blue--Run~UF. The simulations not shown (Runs~UP~and~UQ) behave in a very similar fashion. 

The addition of the extra feedback mechanism (stellar winds) does not alter the evolution of the $\mathcal{Q}$-parameter significantly with respect to the runs with ionisation feedback only, and the results are very similar to those reported in \citet{Parker13a}, namely that star-forming regions which form without feedback lose their structure faster than regions that form with feedback. 

\subsection{Surface densities}
\label{cent_dens}

\citet{Parker13a} noted that the regions that form without feedback lose structure faster than their feedback-influenced counterparts due to their higher initial stellar densities. Without the regulating influence of feedback, regions form with slightly higher initial densities and hence shorter local crossing times, which leads to more interactions and the more rapid loss of substructure. 

The difference in initial density between the runs with and without feedback is readily apparent when examining the median local surface density. We calculate the local stellar surface density following the prescription of \citet{Casertano85}, modified to account for the analysis in projection. For an individual star the local stellar surface density is given by
\begin{equation}
\Sigma = \frac{N - 1} {\pi r_{N}^2},
\label{sigma}
\end{equation}
where $r_{N}$ is the distance to the $N^{\rm th}$ nearest neighbouring star (we adopt $N = 10$ throughout this work; lower $N$ values could bias $\Sigma$ to higher values due to binaries, and higher $N$ values would remove the `localness' from the determination).

\begin{figure}
\begin{center}
\rotatebox{270}{\includegraphics[scale=0.4]{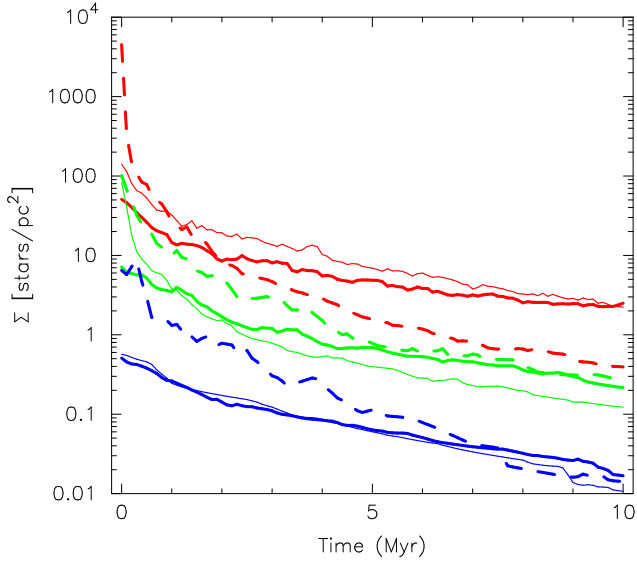}}
\end{center}
\caption[bf]{Evolution of the median stellar surface density for three of the five triplets of simulations (we omit runs UP and UQ from the plot for clarity, but their behaviour follows the same pattern). Simulations \emph{with} feedback are shown by the solid lines; those with ionization feedback only \citep[presented in][]{Parker13a} are shown by the thinner lines, and those with ionization feedback and stellar winds are shown by the thicker lines. Simulations with no feedback are shown by the dashed lines. The red lines are Runs J, green lines are Runs I, dark blue lines are Runs UF. Simulations that formed with feedback all have lower initial densities than those that form without feedback, but the subsequent evolution is non-uniform.}
\label{Dens_all}
\end{figure}

In Fig.~\ref{Dens_all} we show the evolution of the median stellar surface density, $\tilde{\Sigma}_{\rm all}$, again for the evolution of Runs~J, I and UF. When a region is substructured it is usually meaningless to define a `central' or `core' density, and \citet{Parker13a} show that a much better tracer of the true density of a region is the median stellar surface density. The evolution of the runs that form without feedback are shown by the dashed lines, the runs that form with ionisation feedback only are shown by the thin solid lines, and the runs that form with ionisation feedback and stellar winds \citep{Dale14} are shown by the thick lines.

In all sets of simulations, the regions that form without feedback have higher median densities than the simulations that form with feedback. It is these higher densities that facilitate the erasure of substructure, as discussed in \citet{Parker13a} and \citet{Parker14b}. 

\begin{figure}
\begin{center}
\rotatebox{270}{\includegraphics[scale=0.4]{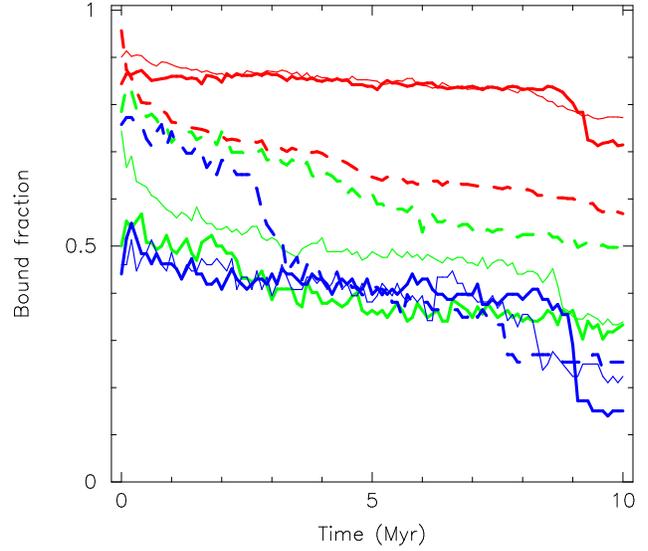}}
\end{center}
\caption[bf]{Evolution of the number fraction of bound stars for three of the five triplets of simulations (we omit runs UP and UQ from the plot for clarity, but their behaviour follows the same pattern as runs UF). Simulations \emph{with} feedback are shown by the solid lines; those with ionization feedback only \citep[presented in][]{Parker13a} are 
shown by the thinner lines, and those with ionization feedback and stellar winds are shown by the thicker lines. Simulations with no feedback are shown by the dashed lines. The red lines are Runs J, green lines are Runs I and the dark blue lines are Runs UF. There is no correlation between the evolution of the fraction of bound stars and the initial conditions (i.e.\,\,feedback versus no feedback).}
\label{bound_3sims}
\end{figure}

\subsection{Bound/unbound stars}

We track the number of stars that remain bound as a function of time by calculating the kinetic and potential energies for each star, as detailed in \citet{Baumgardt02} and \citet{Kruijssen12a}. The potential energy of an individual star, $V_i$, is given by:
\begin{equation}
V_i = - \sum\limits_{i \not= j} \frac{Gm_im_j}{r_{ij}},
\end{equation} 
where $m_i$ and $m_j$ are the masses of two stars and $r_{ij}$ is the distance between them. The kinetic energy of a star, $T_i$ is given thus:
\begin{equation}
T_i = \frac{1}{2}m_i|{\bf v}_i - {\bf v}_{\rm cl}|^2, 
\end{equation}
where ${\bf v}_i$ and ${\bf v}_{\rm cl}$ are the velocity vectors of the star and the centre of mass of the region, respectively. A star is bound if $T_i + V_i < 0$.

In Fig.~\ref{bound_3sims} we show the number fraction of stars that remain bound over the $N$-body integration for three of the five sets of simulations. Again, the simulations that formed with feedback are 
shown by the solid lines; the simulations that include ionisation feedback only are shown by the thinner lines, and the simulations with ionisation and stellar winds are shown by the thicker lines. The simulations that formed without feedback are shown by the dashed lines. The colours correspond to the following simulations; red--Run~J, green--Run~I, dark blue--Run~UF. 

Simulations that form without feedback have a higher fraction of bound stars at the end of the SPH runs (possibly due to the higher total mass), but the subsequent $N$-body evolution does not follow a distinct evolutionary path. For example, in Run~J without feedback, the final bound fraction of $f_{\rm bound, f}$ is lower than for the simulations which include feedback. In the other runs, $f_{\rm bound, f}$ tends to be higher for the simulations without feedback. In the case of Run~J, the high initial stellar density (4518\,stars\,pc$^{-2}$) has led to subsequent dynamical interactions unbinding a large fraction of the stars and hence drastically lowered $f_{\rm bound, f}$ compared to the simulations with feedback.    

\subsection{Mass segregation}

Defining mass segregation has become increasingly difficult due to the many disparate methods which have been promoted in the recent literature. As we will see, different methods which claim to measure mass segregation may actually give very contradictory results. 

Classically, mass segregation is a signature of the onset of energy equipartition in star clusters, whereby the most massive stars have slower velocities and hence sink to the centre \citep{Spitzer69}. In this scenario, the cluster is dynamically old, centrally concentrated and hence has a well-defined radial profile. One can then take different mass bins and compare the density profiles \citep[e.g.][]{Hillenbrand97,Pinfield98}, or look for variations in the slope of the mass function (or luminosity function) with distance from the cluster centre \citep{Carpenter97,deGrijs02,Gouliermis04}. A related method is to quantify the variation of the `Spitzer radius' -- the rms distance of stars in a cluster around the centre of mass -- with luminosity \citep{Gouliermis09}.

These methods all require the definition of the cluster centre. This, and the choice of binning can lead to complications and misinterpretation in the data \citep{Ascenso09}. Furthermore, the two main avenues of massive star formation -- competitive accretion \citep{Zinnecker82,Bonnell98b} and monolithic collapse \citep{McKee03,Krumholz05} -- both predict that the most massive stars should be more centrally located than the lower mass stars; so-called primordial mass segregation. However, as star formation typically occurs in filaments \citep[e.g.][]{Arzoumanian11}, which usually leads to a hierarchical or substructured spatial distribution of stars, then it becomes almost impossible to define the centre of a star-forming region in order to quantify mass segregation.

In order to avoid the need for defining a centre, several methods have been proposed which compare minimum spanning trees of groups of stars \citep{Allison09a} or local surface density around all stars \citep{Maschberger11}. Another recently proposed technique is to use a minimum spanning tree to define groups of stars, and then determine whether the most massive stars are closer to the centre of the group than the average stars \citep{Kirk10,Kirk14} -- the centre is defined as the median position of all stars in the group. We use these three methods to search for (primordial) mass segregation in the SPH simulations from \citet{Dale12a,Dale13,Dale14} and in the subsequent $N$-body evolution (dynamical mass segregation). 

\begin{figure*}
  \begin{center}
\setlength{\subfigcapskip}{10pt}
\hspace*{-0.3cm}
\subfigure[MST of full region]{\label{kirk_omega-a}\rotatebox{270}{\includegraphics[scale=0.4]{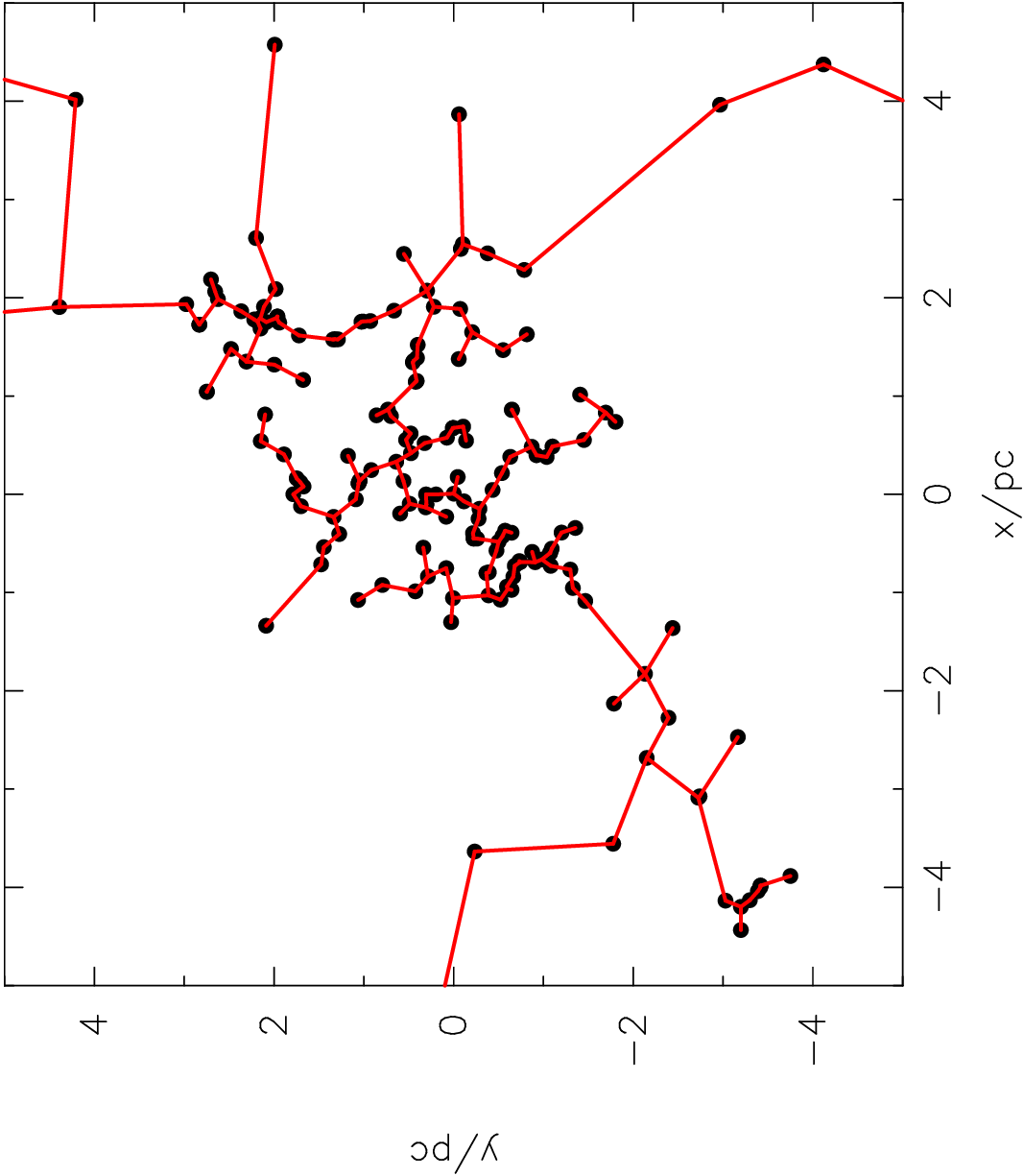}}}  
\hspace*{0.5cm}
\subfigure[Distribution of MST branch lengths]{\label{kirk_omega-b}\rotatebox{270}{\includegraphics[scale=0.4]{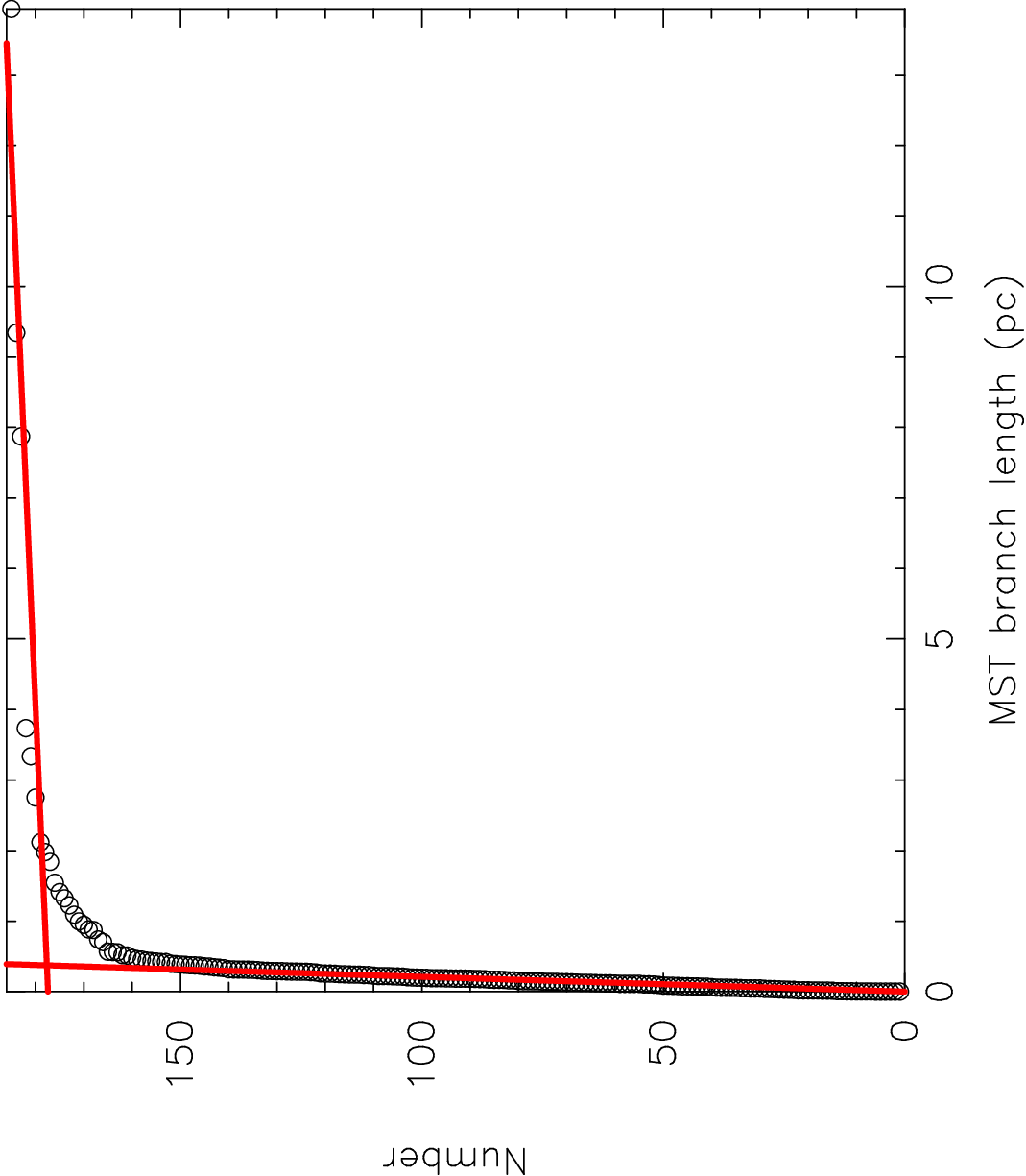}}}
\vspace*{0.25cm}
\hspace*{-0.3cm}
\subfigure[Groups identified by MST break]{\label{kirk_omega-c}\rotatebox{270}{\includegraphics[scale=0.4]{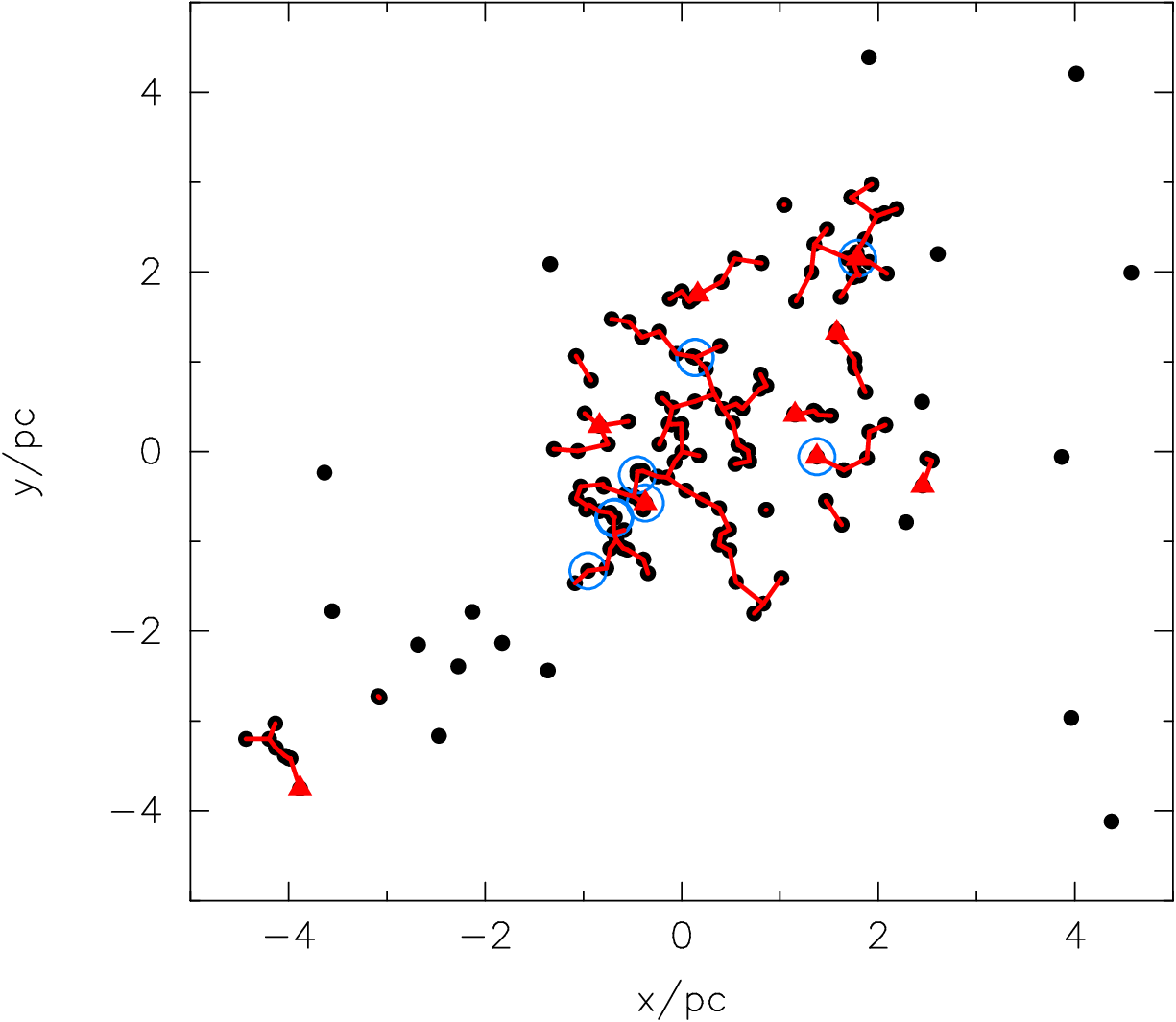}}}  
\hspace*{0.5cm}
\subfigure[Mass ratio to offset ratio for groups]{\label{kirk_omega-d}\rotatebox{270}{\includegraphics[scale=0.4]{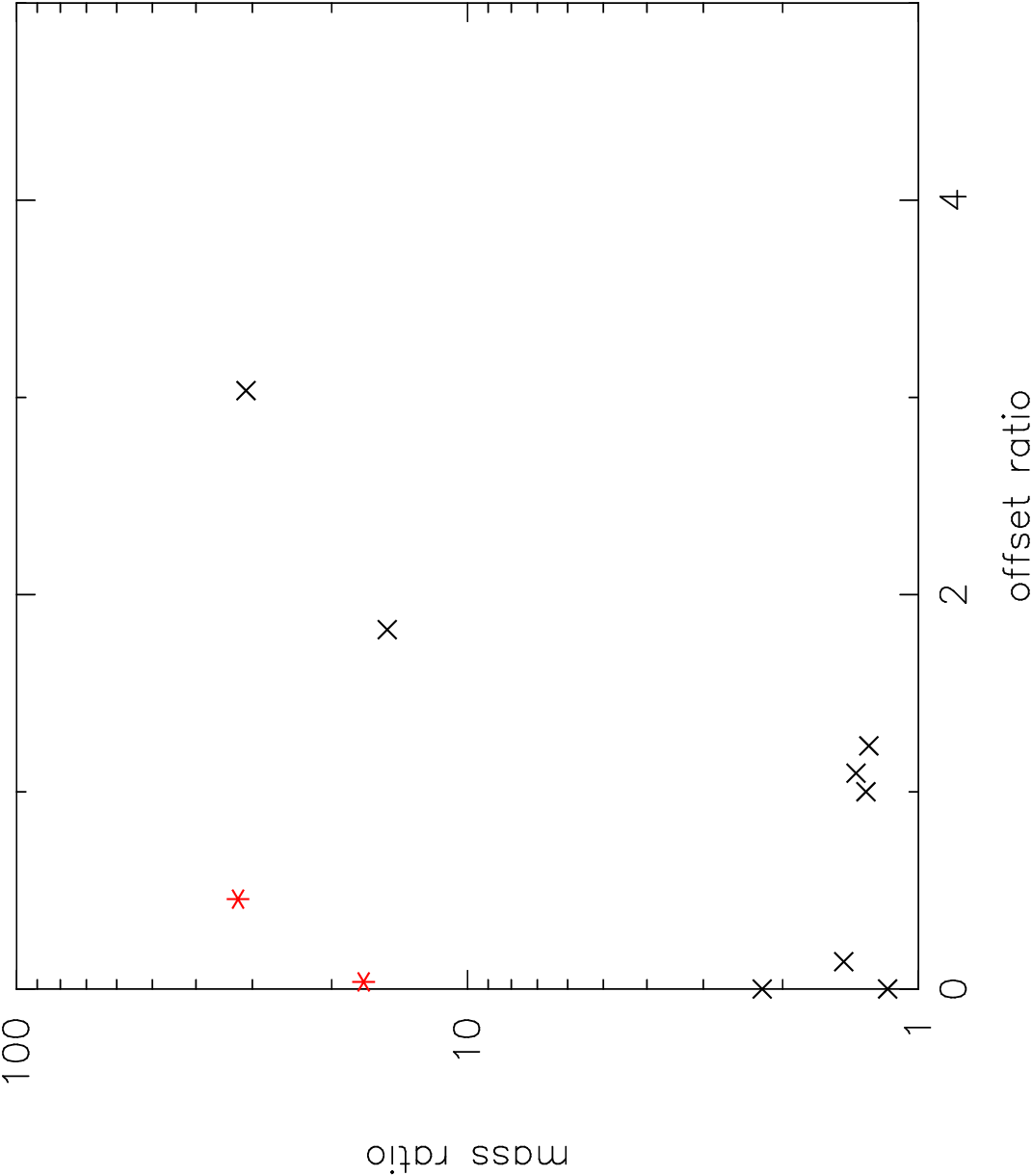}}}
\end{center}
  \caption[bf]{Mass segregation analysis of stellar groups (as in \citet{Kirk10} and \citet{Kirk14}) in the $N$-body simulation of Run~I without feedback at 1\,Myr. In panel (a) the minimum spanning tree (MST) of the entire star-forming region is shown. In panel (b) the distribution 
of MST branch lengths is shown, with the power law fits to the small branches (steep slope) and the long branches (shallow slope) as defined in \citet{Gutermuth09}, \citet{Kirk10} and \citet{Kirk14}. The chosen break length $d_{\rm break}$
(which defines the groups) is the intersection of these two slopes. In panel (c) the MST lengths exceeding $d_{\rm break}$ have been removed, and there are ten groups containing more than two stars (one is outside the field of view in the plot), and two groups containing more than 
ten stars. The ten most massive stars in the entire simulation are shown by the blue circles, and the most massive star in each of the ten groups is shown by a red triangle. In panel (d) for each group we show the ratio of the most massive star to the median stellar mass versus 
the ratio of the distance from the group centre of the most massive star to the group median. The red asterisk symbols are the groups containing ten or more stars.   }
  \label{kirk_omega}
\end{figure*}

\subsubsection{Group segregation ratio, $\Omega_{\rm GSR}$}

First, we use the method of \citet{Kirk10} and \citet{Kirk14}, in which a minimum spanning tree (MST) is constructed for the entire region (see Fig.~\ref{kirk_omega-a}). The cumulative distribution of all MST branch lengths is then made (Fig.~\ref{kirk_omega-b}). Two power-law slopes are then fitted to the shortest lengths, and the longest lengths, and the intersection of these slopes defines the boundary of subclustering, $d_{\rm break}$ \citep{Gutermuth09}. In Fig.~\ref{kirk_omega-c} the MST of the region is shown, but we have omitted the MST lengths greater than the critical length $d_{\rm break}$. 

The location of the most massive star in each group is shown by the red triangle, and we also show the locations of the ten most massive stars for the entire region by the large blue circles. We then follow the method described in \citet{Kirk10,Kirk14} and look for mass 
segregation within the groups. If the position of the most massive star in the group $r_{\rm mm}$ is closer to the central position than the median value for stars, $r_{\rm med}$, the subcluster is said to be mass segregated. This is shown in Fig.~\ref{kirk_omega-d}, where we plot the 
ratio of the highest mass to the median stellar mass in the subcluster against the offset ratio, $r_{\rm mm}/r_{\rm med}$. 

In Fig.~\ref{kirk_omega-d} we also distinguish between groups with $N > 2$ stars, shown by the black crosses, and groups with $N \geq 10$ 
shown by the red asterisks. According to the definition in \citet{Kirk10,Kirk14}, both subclusters with $N \geq 10$ are mass segregated, because their offset ratios are less than unity. We define a `group segregation ratio', $\Omega_{\rm GSR}$, as 
\begin{equation}
\Omega_{\rm GSR} = \frac{N_{\rm seg}}{N_{\rm grp}},
\end{equation}
where $N_{\rm grp}$ are the number of groups, and $N_{\rm seg}$ is the number of these groups that have an offset ratio less than unity. The snapshot shown in Fig.~\ref{kirk_omega} has $\Omega_{\rm GSR} = 1$ for groups with $N \geq 10$, i.e.\,\,all of these groups are mass segregated. 

\subsubsection{Mass segregation ratio, $\Lambda_{\rm MSR}$}

We then use the $\Lambda_{\rm MSR}$ mass segregation ratio pioneered by \citet{Allison09a}. We find the MST of the $N_{\rm MST}$ stars in the chosen subset and
compare this to the MST of sets of $N_{\rm MST}$ random  stars in the
region. If the length of the MST of the chosen subset is shorter than
the average length of the MSTs for the  random stars then the subset
has a more concentrated distribution and is said to be mass segregated. Conversely, if the MST  length of the chosen subset is
longer than the average MST length, then the subset has a less
concentrated distribution, and is  said to be inversely mass
segregated \citep[see e.g.][]{Parker11b}. Alternatively, if the MST length of the chosen subset is
equal to the random MST length,  we can conclude that no mass
segregation is present.

By taking the ratio of the average (mean) random MST length to the subset MST
length, a quantitative measure of the degree of  mass segregation
(normal or inverse) can be obtained. We first determine the subset MST
length, $l_{\rm subset}$. We then  determine the average length of
sets of $N_{\rm MST}$ random stars each time, $\langle l_{\rm average}
\rangle$. There is a dispersion  associated with the average length of
random MSTs, which is roughly Gaussian and can be quantified as the
standard deviation  of the lengths  $\langle l_{\rm average} \rangle
\pm \sigma_{\rm average}$. However, we conservatively estimate the lower (upper) uncertainty 
as the MST length which lies 1/6 (5/6) of the way through an ordered list of all the random lengths (corresponding to a 66 per cent deviation from 
the median value, $\langle l_{\rm average} \rangle$). This determination 
prevents a single outlying object from heavily influencing the uncertainty. 
We can now define the `mass  segregation ratio' 
($\Lambda_{\rm MSR}$) as the ratio between the average random MST pathlength 
and that of a chosen subset, or mass range of objects:
\begin{equation}
\Lambda_{\rm MSR} = {\frac{\langle l_{\rm average} \rangle}{l_{\rm subset}}} ^{+ {\sigma_{\rm 5/6}}/{l_{\rm subset}}}_{- {\sigma_{\rm 1/6}}/{l_{\rm subset}}}.
\end{equation}
$\Lambda_{\rm MSR}$ of $\sim$ 1 shows that the stars in the chosen
subset are distributed in the same way as all the other  stars,
whereas $\Lambda_{\rm MSR} > 1$ indicates mass segregation and
$\Lambda_{\rm MSR} < 1$ indicates inverse mass segregation,
i.e.\,\,the chosen subset is more sparsely distributed than the other stars.

\begin{figure}
\begin{center}
\rotatebox{270}{\includegraphics[scale=0.4]{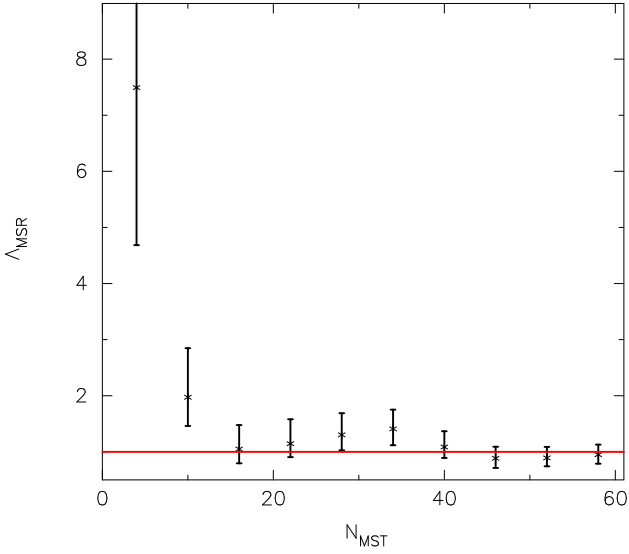}}
\end{center}
\caption[bf]{Mass segregation as defined by $\Lambda_{\rm MSR}$ \citep{Allison09a} in the $N$-body simulation of Run~I without feedback at 1\,Myr. $\Lambda_{\rm MSR} = 1$ (i.e.\,\,no mass segregation) is 
indicated by the red horizontal line.}
\label{lambda}
\end{figure}

In Fig.~\ref{lambda} we show $\Lambda_{\rm MSR}$ as a function of the $N_{\rm MST}$ stars in the subset for Run~I without feedback after 1\,Myr of evolution. The four most massive stars are strongly mass segregated, and this also extends to the 10 most massive stars, although the significance is more marginal.


\subsubsection{Local density ratio, $\Sigma_{\rm LDR}$}

\begin{figure}
\begin{center}
\rotatebox{270}{\includegraphics[scale=0.4]{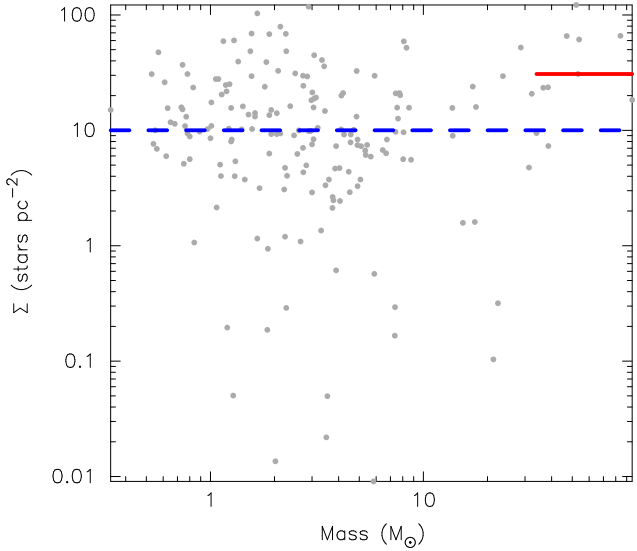}}
\end{center}
\caption[bf]{The $\Sigma - m$ plot \citep{Maschberger11} for the $N$-body simulation of Run~I without feedback at 1\,Myr. The median  surface density in the region is 10\,stars\,pc$^{-2}$ and is shown by the dashed blue line. The 
median surface density of the ten most massive stars is 30\,stars\,pc$^{-2}$, as shown by the solid red line.}
\label{sigm}
\end{figure}

The local surface density of massive stars compared to the median surface density of the full region was pioneered by \citet{Maschberger11} 
as a way of defining mass segregation but minimising the effects of outliers in the distribution. In this definition, the massive stars have no knowledge of each other, but if their surface density distribution can be shown to be inconsistent with the median surface density distribution of the entire region (by means of a KS-test), then the region is said to be mass segregated. In Fig.~\ref{sigm} we show $\Sigma$ versus $m$ for every star for  Run~I without feedback after 1\,Myr of $N$-body evolution.

\citet{Kupper11} and \citet{Parker14b} took 
the ratio of the median surface density of the 10 most massive stars (the red line in Fig.~\ref{sigm}) to the region median (the blue dashed line in   Fig.~\ref{sigm}) to define a `local surface density ratio', $\Sigma_{\rm LDR}$:
\begin{equation}
\Sigma_{\rm LDR} = \frac{\tilde{\Sigma}_\mathrm{subset}}{\tilde{\Sigma}_\mathrm{all}}.
\end{equation}
The  massive stars in this simulation have a higher median surface density than the median value in the region ($\Sigma_{\rm LDR} = 3.08$, and a KS test returns a p-value of $1.35 \times 10^{-2}$ that the two subsets share the same parent distribution).

\subsubsection{Evolution over 10\,Myr}

\begin{figure*}
  \begin{center}
\setlength{\subfigcapskip}{10pt}
\hspace*{-0.3cm}
\subfigure[$\Omega_{\rm GSR}$]{\label{mass_seg_N2-a}\rotatebox{270}{\includegraphics[scale=0.25]{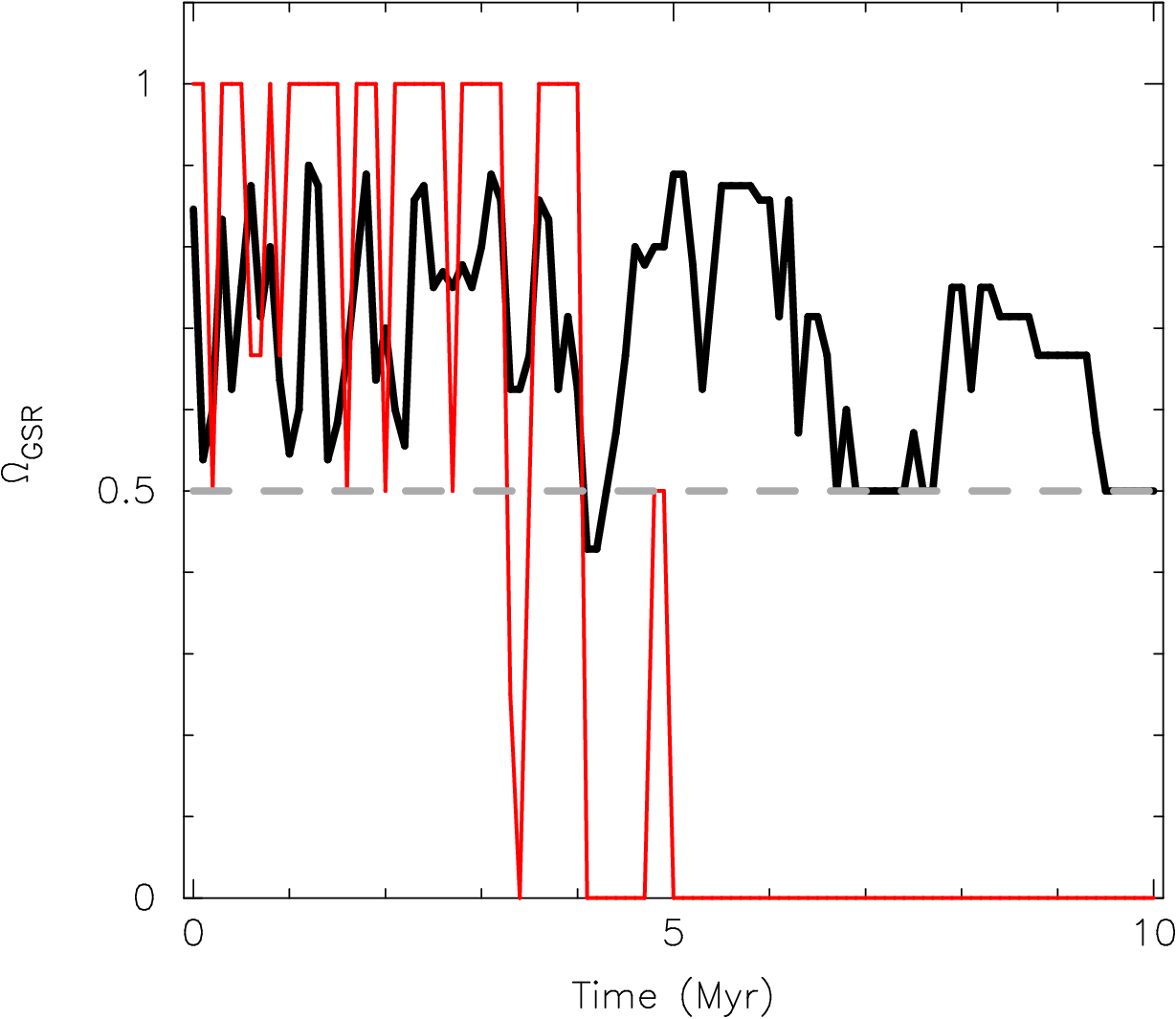}}}  
\hspace*{0.5cm}
\subfigure[$\Lambda_{\rm MSR}$]{\label{mass_seg_N2-b}\rotatebox{270}{\includegraphics[scale=0.25]{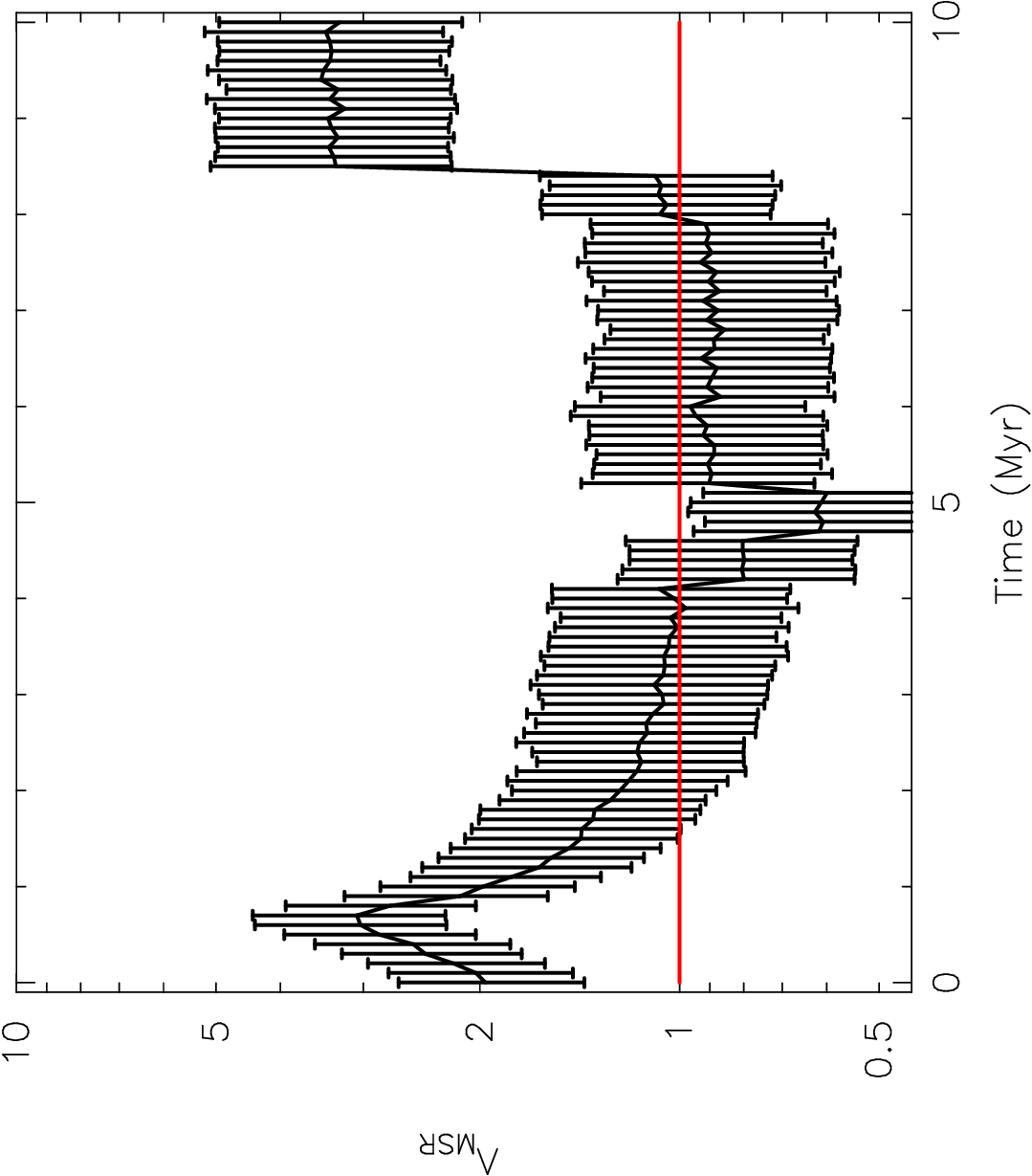}}}  
\hspace*{0.5cm}
\subfigure[$\Sigma_{\rm LDR}$]{\label{mass_seg_N2-c}\rotatebox{270}{\includegraphics[scale=0.25]{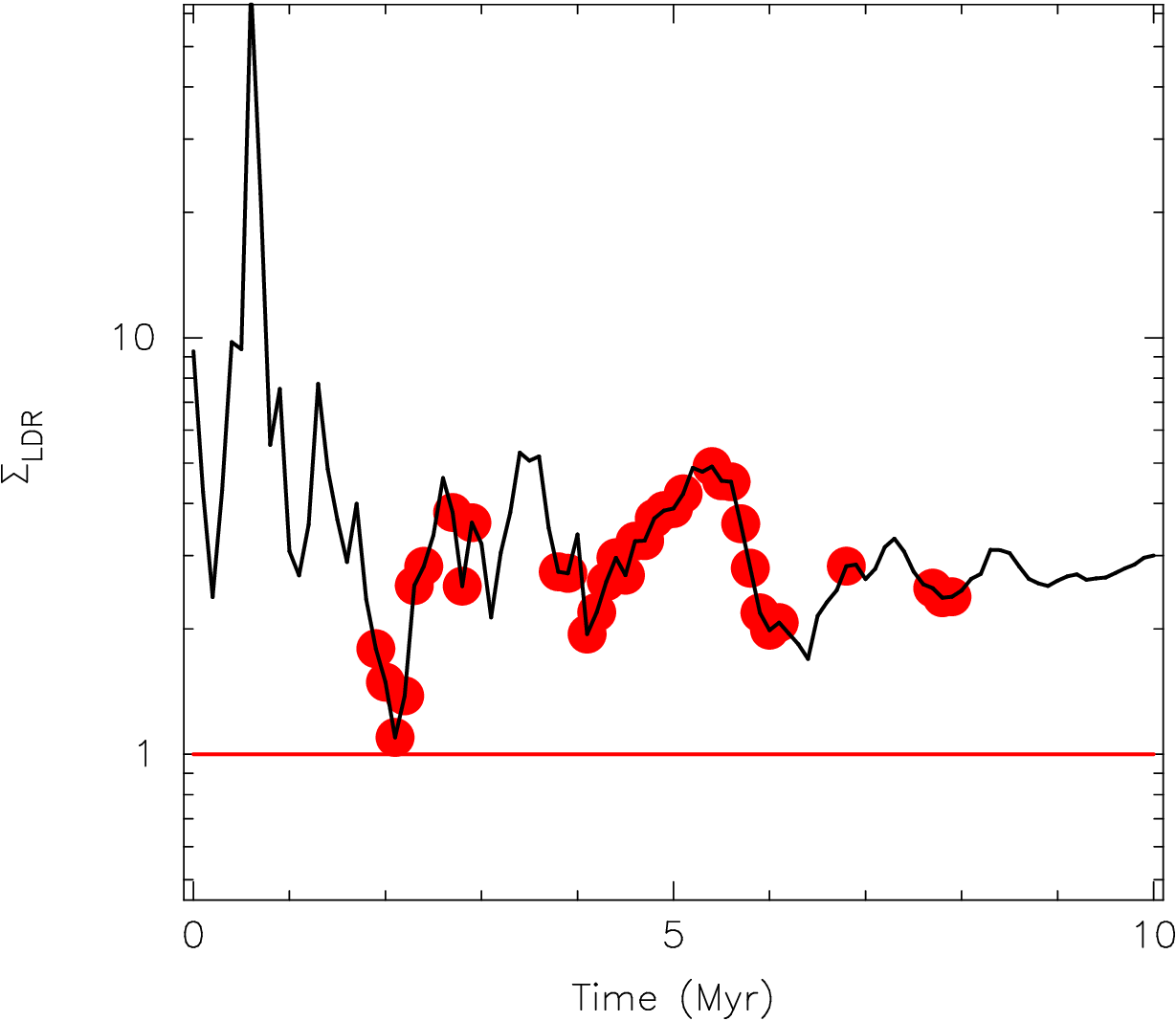}}}
\end{center}
  \caption[bf]{Temporal evolution of three separate measures of mass segregation for the $N$-body simulations of Run~I without feedback;  the fraction of groups identified by $d_{\rm break}$ where the most massive star is nearer the centre of the group 
($\Omega_{\rm GSR}$, panel a) $\Lambda_{\rm MSR}$ (panel b) and $\Sigma_{\rm LDR}$ (panel c) and. In panel (a), we show the fraction of groups 
which have the most massive star closer to the group centre than the median for all groups with more than 2 stars by the thick black line; and the fraction for groups with 10 or more stars by the  thinner red line. The dashed grey line indicates $\Omega_{\rm GSR} = 0.5$, where half of the groups are mass segregated according to this definition. In panel (b) $\Lambda_{\rm MSR}$ is for the 10 most massive stars compared to the region average and $\Lambda_{\rm MSR} = 1$ (i.e.\,\,no preferred spatial distribution) is shown by the solid horizontal red line. In panel (c) $\Sigma_{\rm LDR}$ is also for the 
10 most massive stars compared to the region average and $\Sigma_{\rm LDR} = 1$ (no preferred 
surface density) is shown by the solid horizontal red line. If a KS test between the most massive stars and the full region returns a p-value of more than 0.1 (i.e. the difference is not significant), we plot a solid red circle. }
  \label{mass_seg_N2}
\end{figure*}

We show the evolution of this simulation (Run~I, no feedback) over the full 10\,Myr of $N$-body evolution in Fig.~\ref{mass_seg_N2}. 
According to all three measures, this star-forming region has primordial mass segregation. However, the most massive stars evolve and so the subset 
of the 10 most massive stars in the simulation is not constant. For this reason, the primordial mass segregation disappears according to 
$\Lambda_{\rm MSR}$ and $\Sigma_{\rm LDR}$, until dynamical evolution causes a ``re-segregation'' after 8\,Myr. This is not apparent from the $\Omega_{\rm GSR}$, which is heavily dependent on the group definition, rather than the locations of the most massive stars.

\begin{figure*}
  \begin{center}
\setlength{\subfigcapskip}{10pt}
\hspace*{-0.3cm}
\subfigure[$\Omega_{\rm GSR}$]{\label{mass_seg_B2-a}\rotatebox{270}{\includegraphics[scale=0.25]{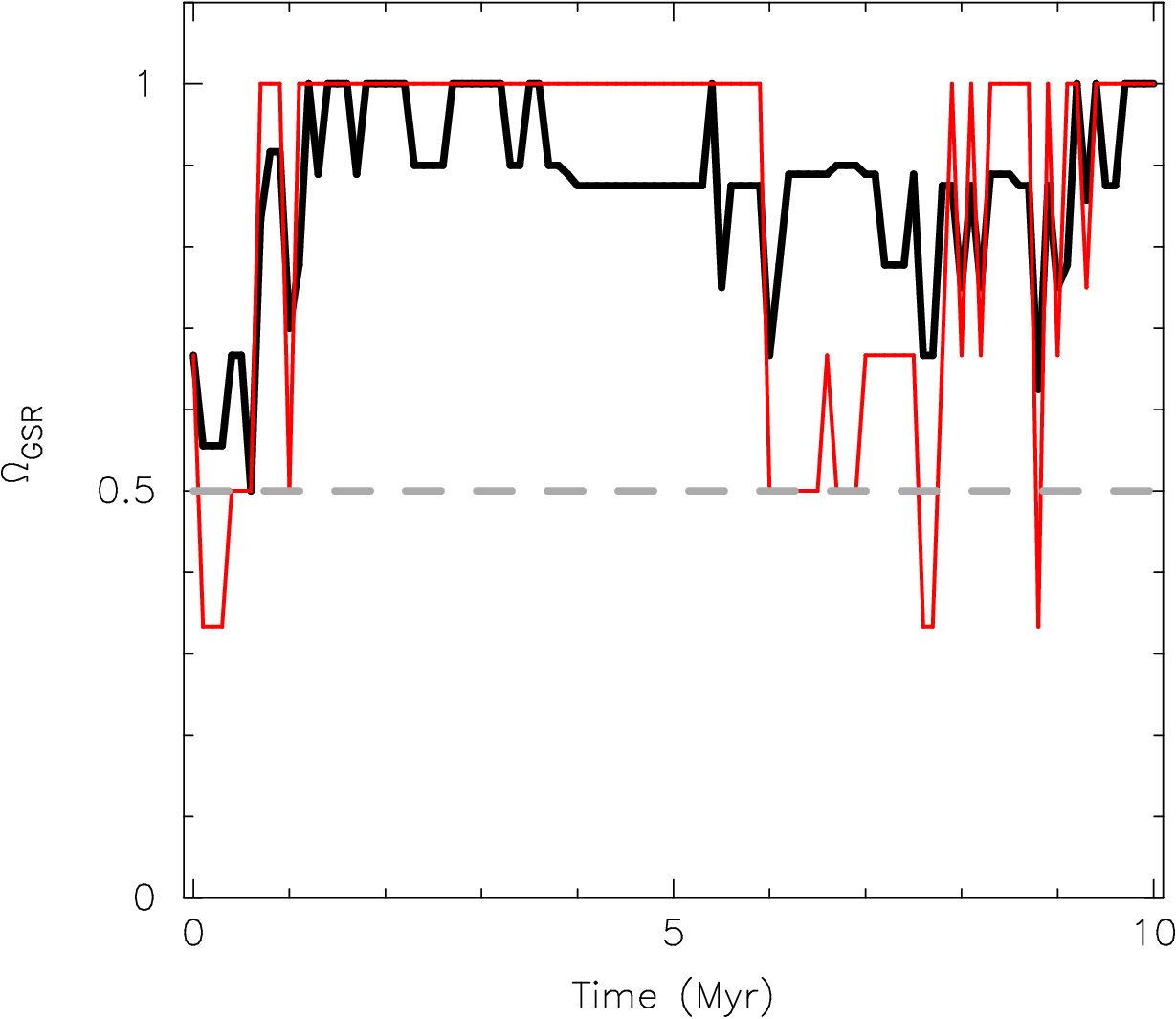}}} 
\hspace*{0.5cm} 
\subfigure[$\Lambda_{\rm MSR}$]{\label{mass_seg_B2-b}\rotatebox{270}{\includegraphics[scale=0.25]{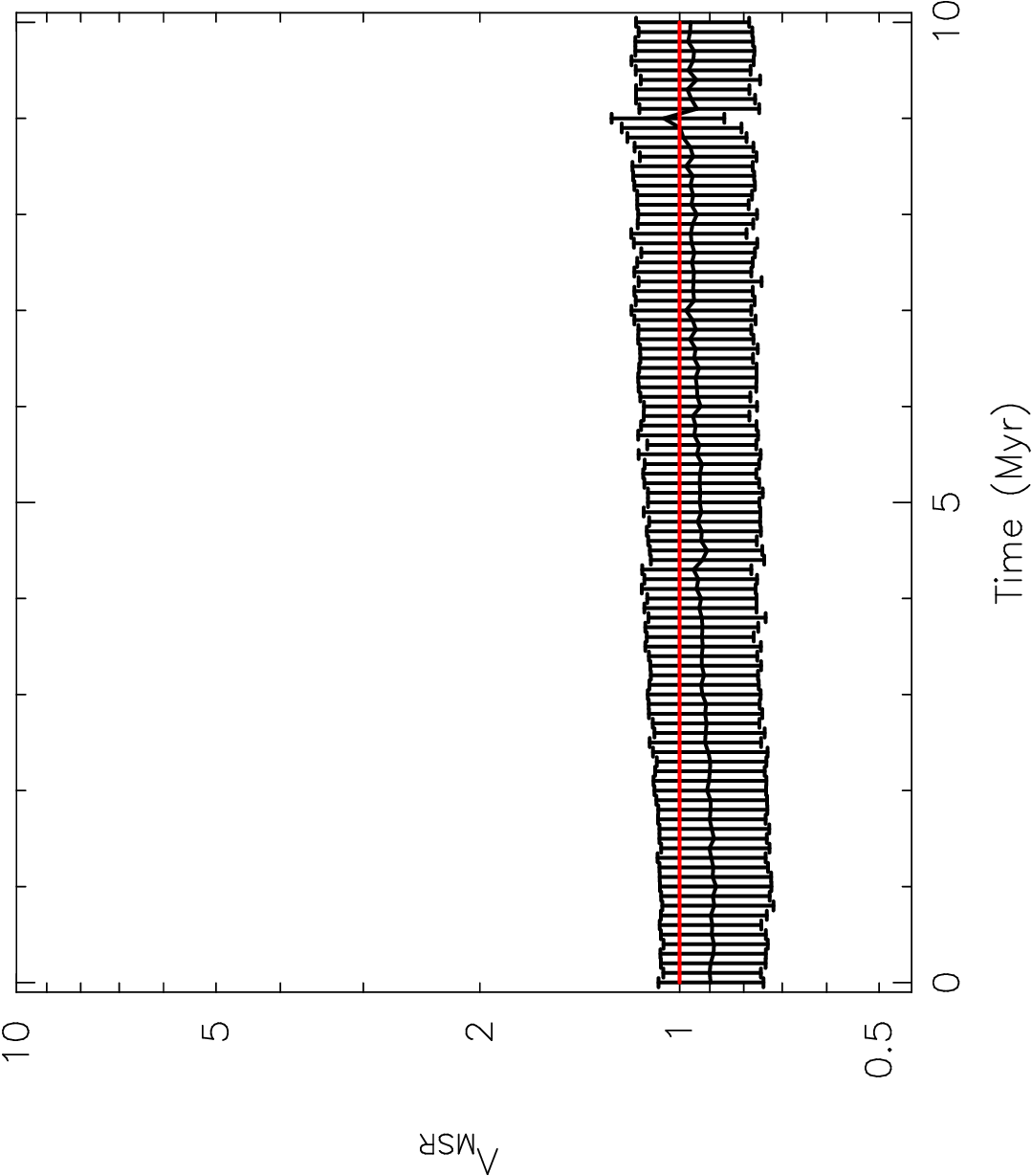}}}  
\hspace*{0.5cm}
\subfigure[$\Sigma_{\rm LDR}$]{\label{mass_seg_B2-c}\rotatebox{270}{\includegraphics[scale=0.25]{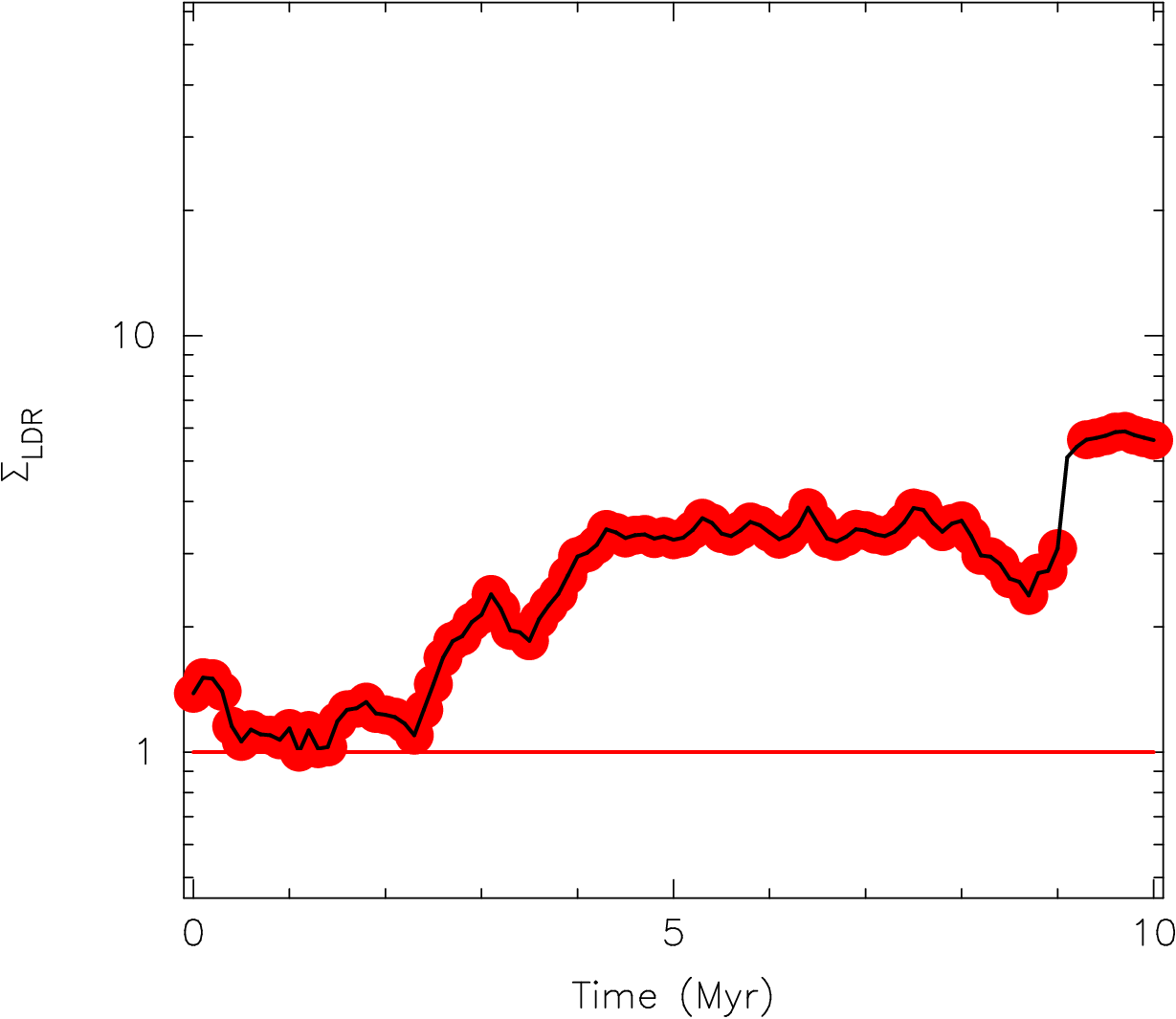}}}
\end{center}
  \caption[bf]{Temporal evolution of three separate measures of mass segregation for the $N$-body simulations of Run~I with both types of feedback; the fraction of groups identified by $d_{\rm break}$ where the most massive star is nearer the centre of the group 
($\Omega_{\rm GSR}$, panel a), $\Lambda_{\rm MSR}$ (panel b) and $\Sigma_{\rm LDR}$ (panel c). In panel (a), we show the fraction of groups 
which have the most massive star closer to the group centre than the median for all groups with more than 2 stars by the thick black line; and the fraction for groups with 10 or more stars by the  thinner red line. The dashed grey line indicates $\Omega_{\rm GSR} = 0.5$, where half of the groups are mass segregated according to this definition. In panel (b) $\Lambda_{\rm MSR}$ is for the 10 most massive stars compared to the region average and $\Lambda_{\rm MSR} = 1$ (i.e.\,\,no preferred spatial distribution) is shown by the solid horizontal red line. In panel (c) $\Sigma_{\rm LDR}$ is also for the 
10 most massive stars compared to the region average and $\Sigma_{\rm LDR} = 1$ (no preferred 
surface density) is shown by the solid horizontal red line. If a KS test between the most massive stars and the full region returns a p-value of more than 0.1 (i.e. the difference is not significant), we plot a solid red circle. }
  \label{mass_seg_B2}
\end{figure*}

We then show the evolution of the same SPH simulation, but this time with both ionisation and wind feedback in Fig.~\ref{mass_seg_B2}. In this simulation there is no significant mass segregation according to $\Lambda_{\rm MSR}$ or $\Sigma_{\rm LDR}$, whereas $\Omega_{\rm GSR}$ does suggest the star-forming region is mass segregated.

\subsection{Structure versus mass segregation}

For a large set of purely $N$-body simulations \citet{Parker14b} showed that plotting spatial structure against mass segregation measurements can distinguish between the initial conditions of star-forming regions. In the following section we show $\mathcal{Q} - \Omega_{\rm GSR}$, $\mathcal{Q} - \Lambda_{\rm MSR}$, and then $\mathcal{Q} - \Sigma_{\rm LDR}$. 

\subsubsection{$\mathcal{Q} - \Omega_{\rm GSR}$}

\begin{figure*}
  \begin{center}
\setlength{\subfigcapskip}{10pt}
\hspace*{-0.3cm}
\subfigure[$\mathcal{Q} - \Omega_{\rm GSR}$, no feedback]{\label{omega_Q-a}\rotatebox{270}{\includegraphics[scale=0.35]{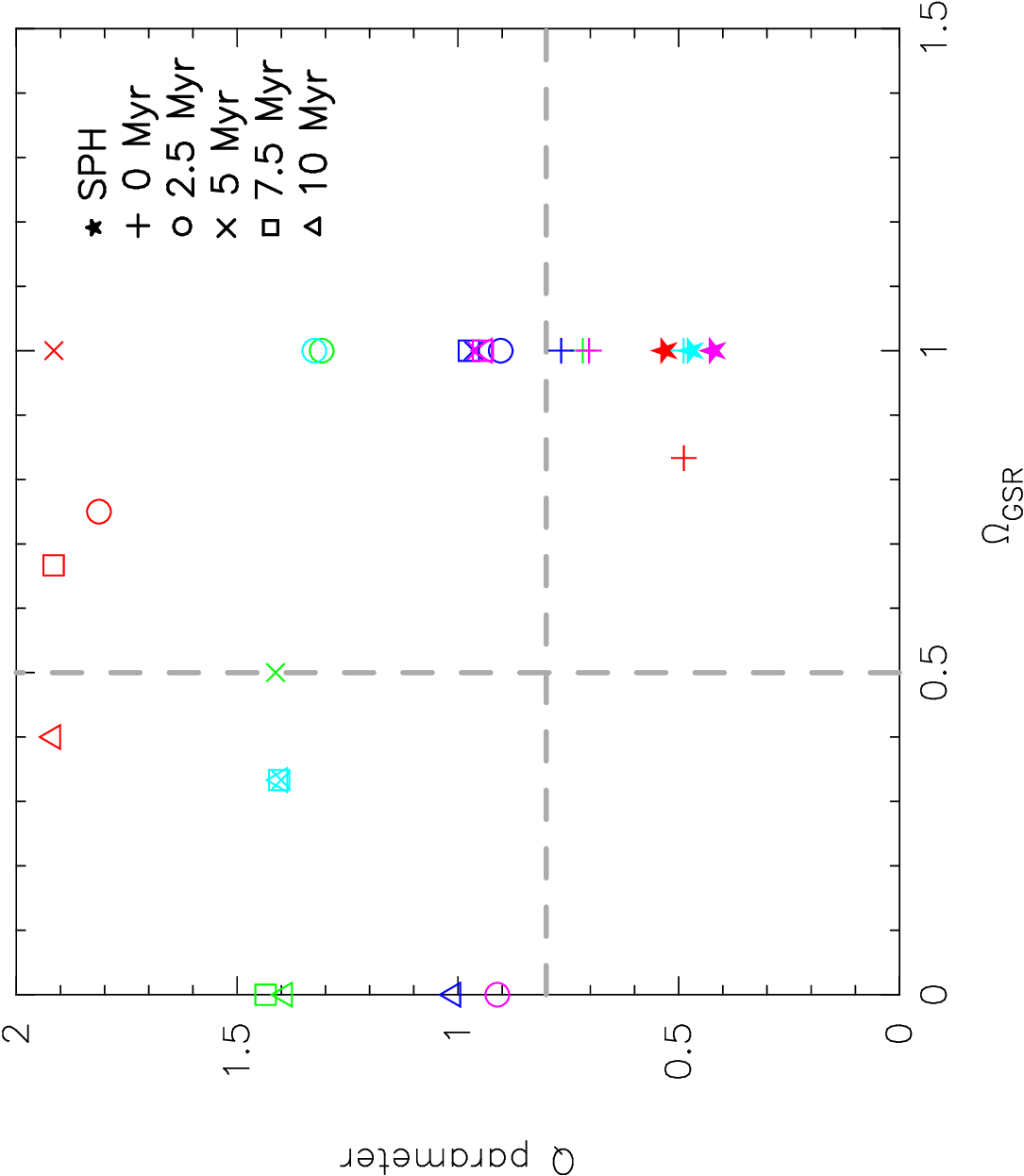}}}  
\hspace*{0.5cm}
\subfigure[$\mathcal{Q} - \Omega_{\rm GSR}$, with feedback]{\label{omega_Q-b}\rotatebox{270}{\includegraphics[scale=0.35]{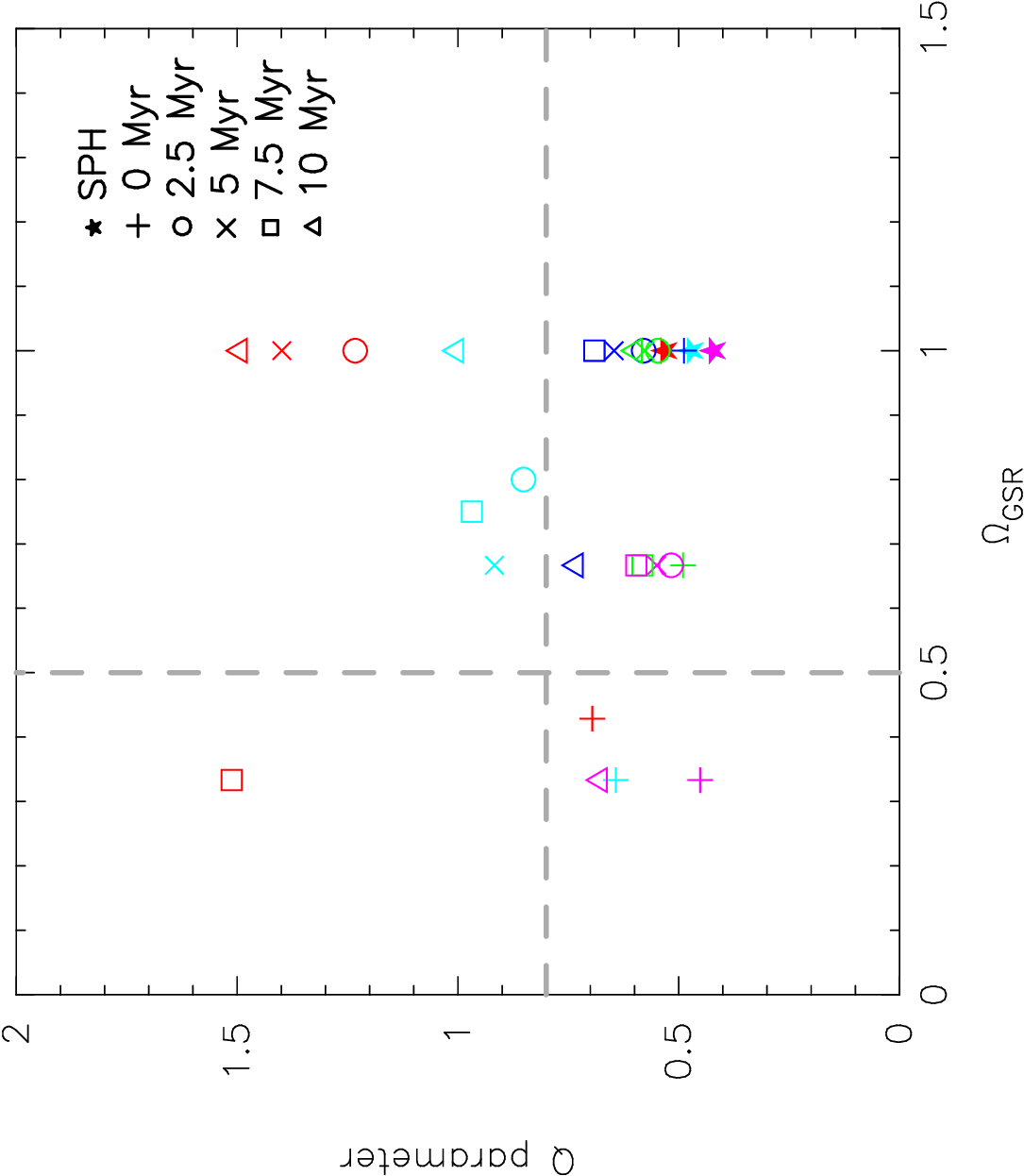}}}
\end{center}
  \caption[bf]{Spatial structure $\mathcal{Q}$ versus $\Omega_{\rm GSR}$. In panel (a) we show the evolution of all five star-forming regions which form without feedback and in panel (b) we show the evolution of all five regions that form with ionisation and stellar wind feedback. The red symbols are Runs J, green symbols are Runs I, dark blue symbols are Runs UF, cyan symbols are Runs UP and the magenta symbols are Runs UQ. In addition, we show the measurements from the SPH simulations before feedback is switched on, apart from Run UF, which does not form enough stars to define distinct subgroups with $N > 10$. Also, Runs I and UQ both have $\mathcal{Q} = 0.42$ and $\Omega_{\rm GSR} = 1$, so only the cyan star symbol is visible in the plot. $\Omega_{\rm GSR} = 0.5$ (half of the groups are mass segregated) is shown by the vertical dashed line, and the boundary between a substructured and centrally concentrated spatial distribution ($\mathcal{Q} = 0.8$) is shown by the horizontal dashed line. }
  \label{omega_Q}
\end{figure*}

 In Fig.~\ref{omega_Q} we show the $\mathcal{Q}$ parameter plotted against the group mass segregation ratio $\Omega_{\rm GSR}$ for the simulations without feedback (Fig.~\ref{omega_Q-a}) and those with ionisation and stellar winds (Fig.~\ref{omega_Q-b}). We show the data at 0, 2.5, 5, 7.5 and 10\,Myr, and we also plot the values from the SPH runs before feedback is switched on, apart from Run UF, which does not contain enough stars at that stage for there to be a group containing ten or more stars. The colour scheme is as follows; red -- Runs J, green -- Runs I, dark blue -- Runs UF, cyan -- Runs UP and magenta -- Runs UQ. (Note that Runs I and UQ both have $\mathcal{Q} = 0.42$ and $\Omega_{\rm GSR} = 1$, so only the cyan star symbol is visible in the plot.)

In this plot,  $\Omega_{\rm GSR} = 0.5$ (half of the groups are mass segregated) is shown by the vertical dashed line, and the boundary between a substructured and centrally concentrated spatial distribution ($\mathcal{Q} = 0.8$) is shown by the horizontal dashed line. Before the feedback mechanisms are switched on, all the groups have their most massive member more centrally concentrated than the average star. When no feedback is switched on (Fig.~\ref{omega_Q-a}), this behaviour carries forward to the end of the SPH simulation ($t = 0$\,Myr in the $N$-body integration). As the regions (and the massive stars within them) evolve, the mass segregation is gradually lost -- so much so that after 10\,Myr only one simulation (UQ -- the magenta triangle) has  $\Omega_{\rm GSR} > 0.5$. 

Conversely, in the simulations with photoionisation and stellar winds, feedback has wiped out the early primordial mass segregation in groups in three out of five simulations (the red (Run J), cyan (Run UP) and magenta (Run UQ) plus signs). However, the subsequent evolution of these regions leads to most simulations having $\Omega_{\rm GSR} > 0.5$ -- i.e.\,\,most of the groups are segregated in the sense that their most massive member is closer to the group centre than the average star. 

\begin{figure*}
  \begin{center}
\setlength{\subfigcapskip}{10pt}
\hspace*{-0.3cm}
\subfigure[$\mathcal{Q} - \Lambda_{\rm MSR}$, no feedback]{\label{lambda_Q-a}\rotatebox{270}{\includegraphics[scale=0.35]{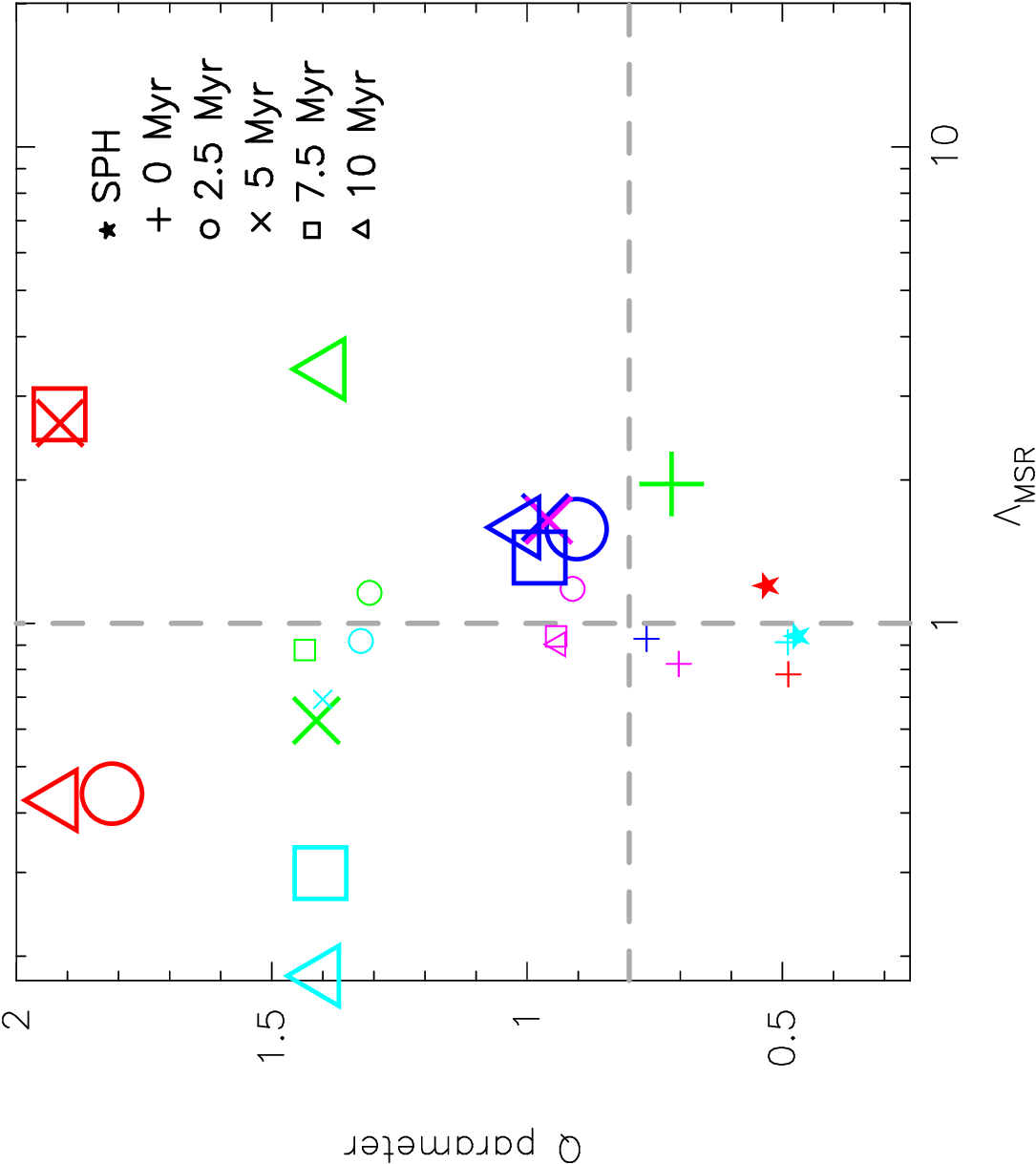}}}  
\hspace*{0.5cm}
\subfigure[$\mathcal{Q} - \Lambda_{\rm MSR}$, with feedback]{\label{lambda_Q-b}\rotatebox{270}{\includegraphics[scale=0.35]{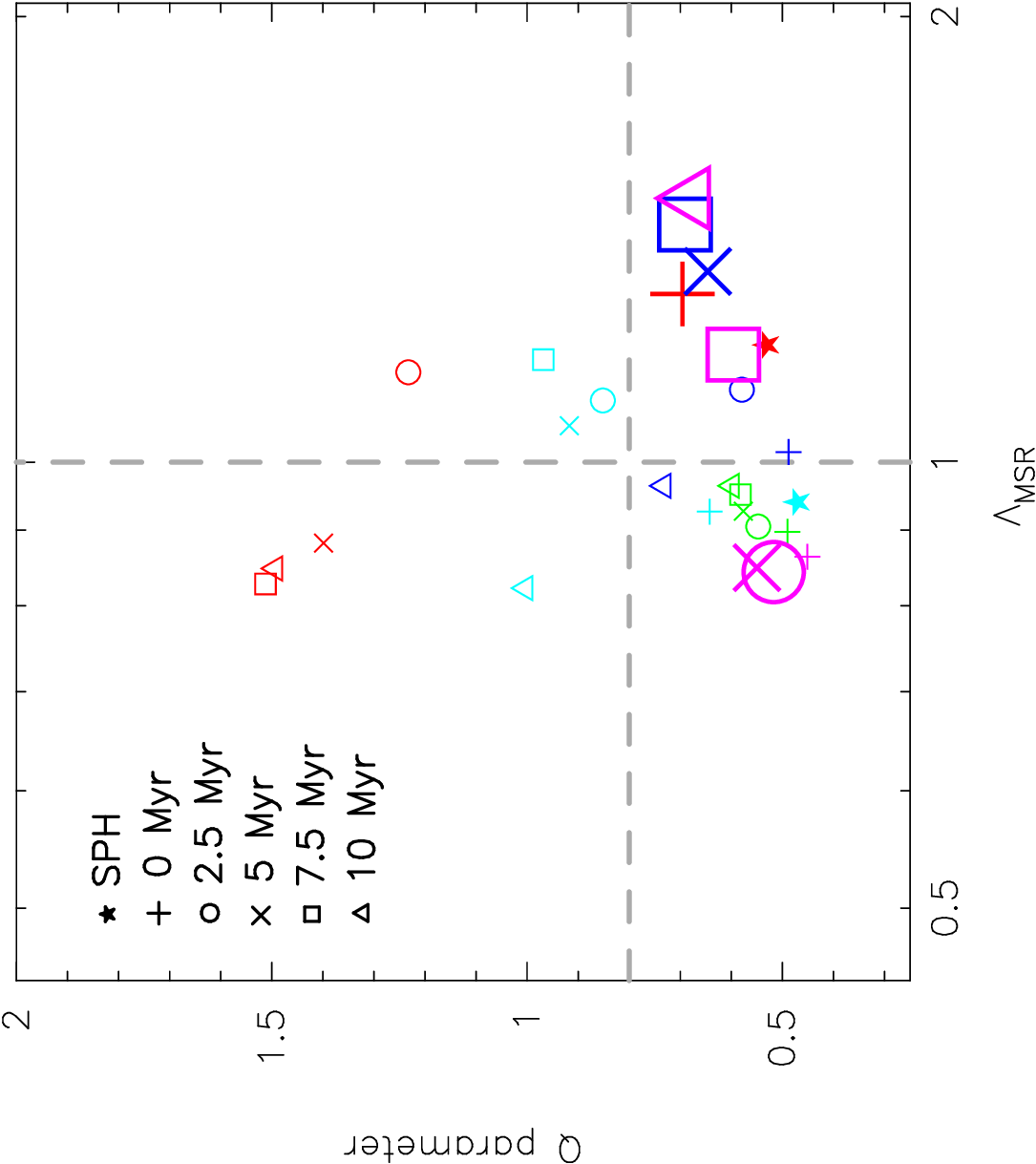}}}
\end{center}
  \caption[bf]{Spatial structure $\mathcal{Q}$ versus $\Lambda_{\rm MSR}$. In panel (a) we show the evolution of all five star-forming regions which form without feedback and in panel (b) we show the evolution of all five regions that form with ionisation and stellar wind feedback. The red symbols are Runs J, green symbols are Runs I, dark blue symbols are Runs UF, cyan symbols are Runs UP and the magenta symbols are Runs UQ. In addition, we show the measurements from the SPH simulations before feedback is switched on (only Runs J and UP have enough stars at that stage to calculate $\Lambda_{\rm MSR}$). The larger symbols indicate where the deviation of $\Lambda_{\rm MSR}$ from unity is significant. $\Lambda_{\rm MSR} = 1$ (no mass segregation) is shown by the vertical dashed line, and the boundary between a substructured and centrally concentrated spatial distribution ($\mathcal{Q} = 0.8$) is shown by the horizontal dashed line. }
  \label{lambda_Q}
\end{figure*}

\subsubsection{$\mathcal{Q} - \Lambda_{\rm MSR}$}

 In Fig.~\ref{lambda_Q} we show the $\mathcal{Q}$ parameter plotted against the mass segregation ratio $\Lambda_{\rm MSR}$ for the simulations without feedback (Fig.~\ref{lambda_Q-a}) and those with ionisation and stellar winds (Fig.~\ref{lambda_Q-b}).

As in Fig.~\ref{omega_Q}, we show the data at 0, 2.5, 5, 7.5 and 10\,Myr, and we also plot the values from the SPH runs before feedback is switched on. At this point in the SPH run, the number of stars that have already formed can be rather low ($<$50) and we only plot the SPH measurement for Runs~J and UP, which already have enough stars to make the determination of $\Lambda_{\rm MSR}$ meaningful. The colour scheme is as follows; red -- Runs J, green -- Runs I, dark blue -- Runs UF, cyan -- Runs UP and magenta -- Runs UQ. The large symbols indicate 
when $\Lambda_{\rm MSR}$ deviates significantly from unity (i.e.\,\,mass segregation, or inverse mass segregation is present). 

The example shown in Fig.~\ref{mass_seg_N2-a} -- Run I without feedback -- is shown by the green symbols in Fig.~\ref{lambda_Q-a}, and the corresponding run with feedback (Fig.~\ref{mass_seg_B2-a}) is shown by the green symbols in Fig.~\ref{lambda_Q-b}. 

Primordial mass segregation as measured by $\Lambda_{\rm MSR}$ is found in one simulation (Run~I) out of five for the regions without feedback, and in one of five simulations (Run~J) with feedback (though not the same simulation). Run~J was not mass segregated before feedback was switched on, with $\Lambda_{\rm MSR} = 1.2$ -- the red star in Fig.~\ref{lambda_Q-a}.  

Considering the runs that formed without feedback, in addition to the behaviour of Run~I which was shown in Fig.~\ref{mass_seg_N2-a}, over 10\,Myr of subsequent dynamical evolution Run~J (the red symbols) fluctuates between being significantly mass segregated (at 5 and 7.5\,Myr), and significantly \emph{inversely} mass segregated (at 2.5 and 10\,Myr). Run UF (the dark blue symbols) dynamically mass segregates and remains so over the full 10\,Myr. Run UP (the cyan points) dynamically inversely mass segregates after 2.5\,Myr and Run UQ (the magenta points) does not significantly mass segregate, or inverse mass segregate aside from a brief snapshot at 5\,Myr.

The runs that form with feedback in general display no primordial mass segregation. The one simulation that does show primordial mass segregation (Run~J) unsegregates, due to stellar evolution -- the ten most massive stars at $t = 0$\,Myr are not the same ten most massive stars even after only 2.5\,Myr. Run UF (the dark blue symbols) dynamically mass segregates until after 7.5\,Myr, when dynamical interactions eject two massive stars and the cluster reverts to being unsegregated after 10\,Myr. Run UP first becomes inversely mass segregated, before dynamical interactions lead to normal mass segregation. 

\begin{figure*}
  \begin{center}
\setlength{\subfigcapskip}{10pt}
\hspace*{-0.3cm}
\subfigure[$\mathcal{Q} - \Sigma_{\rm LDR}$, no feedback]{\label{sigma_Q-a}\rotatebox{270}{\includegraphics[scale=0.35]{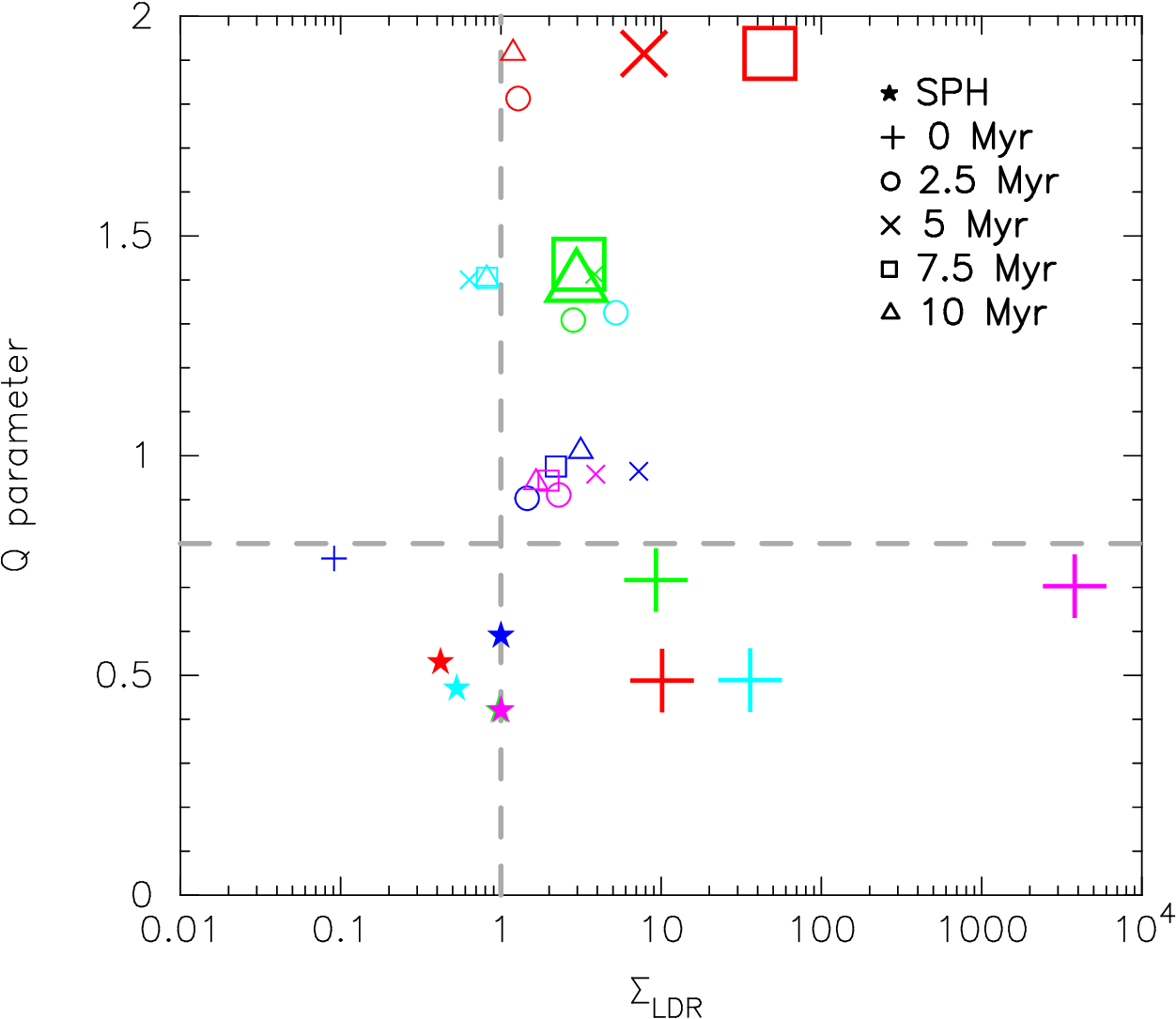}}}  
\hspace*{0.5cm}
\subfigure[$\mathcal{Q} - \Sigma_{\rm LDR}$, with feedback]{\label{sigma_Q-b}\rotatebox{270}{\includegraphics[scale=0.35]{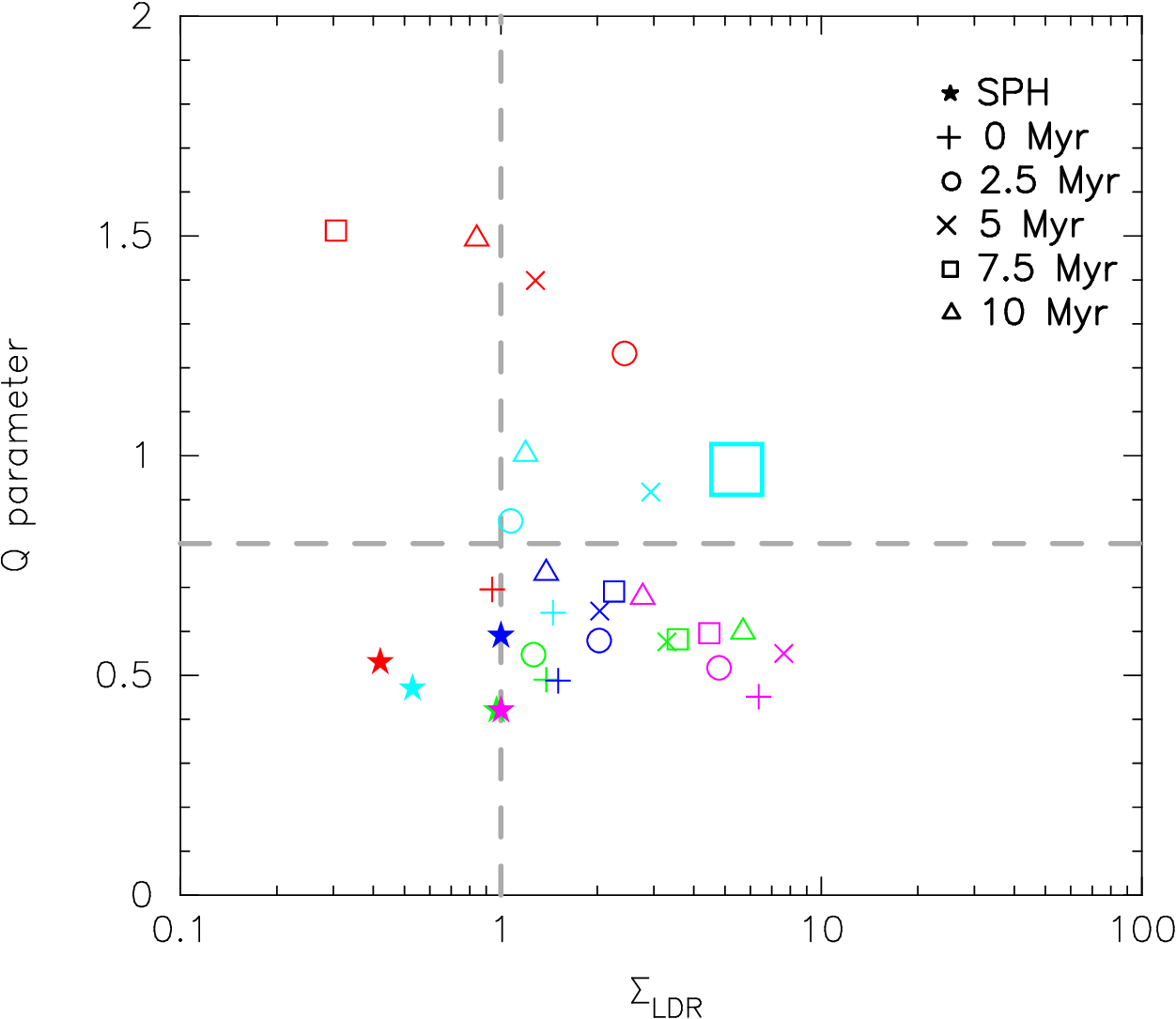}}}
\end{center}
  \caption[bf]{Spatial structure $\mathcal{Q}$ versus $\Sigma_{\rm LDR}$. In panel (a) we show the evolution of all five star-forming regions which form without feedback and in panel (b) we show the evolution of all five regions that form with ionisation and stellar wind feedback. The red symbols are Runs J, green symbols are Runs I, dark blue symbols are Runs UF, cyan symbols are Runs UP and the magenta symbols are Runs UQ. In addition, we show the measurements from the SPH simulations before feedback is switched on. The larger symbols indicate where the deviation of $\Sigma_{\rm LDR}$ from unity is significant. $\Sigma_{\rm LDR} = 1$ (no mass segregation) is shown by the vertical dashed line, and the boundary between a substructured and centrally concentrated spatial distribution ($\mathcal{Q} = 0.8$) is shown by the horizontal dashed line. }
  \label{sigma_Q}
\end{figure*}

\subsubsection{$\mathcal{Q} - \Sigma_{\rm LDR}$}

In Fig.~\ref{sigma_Q} we show the $\mathcal{Q}$ parameter plotted against the local surface density ratio $\Sigma_{\rm LDR}$. Notably, in the simulations without feedback (Fig.~\ref{sigma_Q-a}), nearly \emph{all} simulations show primordial mass segregation at $t = 0$\,Myr, whereas none are mass segregated before feedback is switched on in the SPH runs (the star symbols clustered around $\Sigma_{\rm LDR} = 1$ in both panels). The subsequent combination of dynamical and stellar evolution generally erases this signature, apart from several snapshots in Run J (the red symbols) and Run I (the green symbols). 

When feedback is included, no simulations have primordial mass segregation according to $\Sigma_{\rm LDR}$. Furthermore, following dynamical evolution only one simulation displays a $\Sigma_{\rm LDR}$ ratio that is significantly higher than unity -- Run UP at 7.5\,Myr. 

The large fraction of simulations without feedback that display a high  $\Sigma_{\rm LDR}$ at the end of the SPH calculation compared to the simulations with feedback suggests that the absence of feedback leads to higher stellar densities and hence enables the massive stars to acquire a retinue of lower mass stars \citep[as seen in pure $N$-body simulations,][]{Parker14b}. However, once stellar evolution is included, the massive stars lose so much mass \citep{Parker13a} that the determination of  $\Sigma_{\rm LDR}$ begins to include stars with local densities comparable to the median value in the region.

\section{Discussion}
\label{discuss}

We have expanded upon the work in \citet{Parker13a} and followed the dynamical evolution of five SPH simulations of star formation which also include stellar winds, as well as photoionisation feedback, and compare them to control run simulations which have feedback switched off. The inclusion of stellar winds does not appreciably affect the formation of a region any more so than in the simulations which only include photoionisation as a feedback source. The results are similar to those in \citet{Parker13a}; regions which form under the influence of feedback have lower stellar densities than those that do not. The systematically lower densities raise the local crossing time in the regions, and this leads to the regions retaining substructure far longer than the regions which are not influenced by feedback. 

We also look for mass segregation in both the non-feedback, and the feedback influenced star-forming regions. Recent analyses of hydrodynamical simulations of star formation by \citet{Kirk14} and \citet{Myers14} have claimed to find primordial mass segregation; the massive stars segregate early on in the calculation and remain segregated throughout. 

As discussed in \citet{Parker14b} and Parker \& Goodwin, in prep., different methods to search for mass segregation routinely produce apparently contradictory results, especially for complex spatial distributions. For example, the $\Lambda_{\rm MSR}$ ratio \citep{Allison09a} measures the spatial distribution of a subset of massive stars compared to random subsets, whereas the $\Sigma_{\rm LDR}$ ratio \citep{Maschberger11,Kupper11,Parker14b} measures the surface density around individual massive stars, and compares this to the surface density around average stars. Finally, the method from  \citet{Kirk10,Kirk14}, which we use to define a `group segregation ratio' -- $\Omega_{\rm GSR}$,  measures the distance of the most massive star from the centre of a subcluster to the distance of the median mass star from the centre. 

$\Lambda_{\rm MSR}$ provides information on the \emph{global} spatial distribution of massive stars, whereas $\Sigma_{\rm LDR}$ provides information on the local density around those stars. As shown in Parker \& Goodwin, in prep., the $\Omega_{\rm GSR}$ method sometimes reflects the local density of massive stars, but is reliant on dividing up a star-forming region into individual groups. It is not clear whether this assumption is valid; for example, in these simulations there is only one star formation `event' and the massive stars -- and associated low mass stars -- are not independent of each other.

This is apparent in Figs.~\ref{mass_seg_N2}~and~\ref{mass_seg_B2}, where $\Lambda_{\rm MSR}$ and $\Sigma_{\rm LDR}$ display similar behaviour, whereas $\Omega_{\rm GSR}$ does not. More often than not, $\Omega_{\rm GSR}$ shows a region to be mass segregated on local scales when the global distribution is consistent with a random distribution. We therefore suggest that the claim of primordial mass segregation in clusters by \citet{Kirk14} has been influenced by the method used to define mass segregation and does not reflect the true spatial distribution from the outcome of star formation.

Furthermore, when we examine the results of several SPH simulations, we see primordial mass segregation in some, but not all, simulations (either four in five, or two in five using $\Omega_{\rm GSR}$, one in five using $\Lambda_{\rm MSR}$, and either none, or four in five using $\Sigma_{\rm LDR}$ depending on whether feedback was switched on or not). Interestingly, \citet{Myers14} find mass segregation according to $\Sigma_{\rm LDR}$ in their magneto-hydrodynamical simulations of star formation which include feedback, whereas we only see high $\Sigma_{\rm LDR}$ ratios in the simulations without feedback. 

In order to compare our results to those of \citeauthor{Myers14}, we estimate the $\Sigma_{\rm LDR}$ ratio from their fig.~17\footnote{We believe there is an error in figs.~16~and~17 in \citet{Myers14}, in that the simulation run with the `strong' magnetic field in their work is symbolled and labelled as their control run simulation with no magnetic field, and vice versa.} using their median cluster values (the coloured lines in their fig.~17) and the median of the 10 most massive stars as shown in their figure (using all stars with masses $>$1\,M$_\odot$ gives almost identical results). In their control (hydro only) run $\Sigma_{\rm LDR} = 1.4$, in their run with a `Weak' magnetic field ($\Sigma_{\rm LDR} = 4.4$) and in their run with a `Strong' magnetic field $\Sigma_{\rm LDR} = 4.4$. The presence of a magnetic field may have caused mass segregation in this particular simulation, but the level of mass segregation according to $\Sigma_{\rm LDR}$ does not increase with increasing magnetic field. As we have seen with the simulations from  \citet{Dale12a,Dale13,Dale14} mass segregation can be random, and more than one simulation is required to ascertain whether the addition of extra physics is the root cause of a different spatial distribution for the most massive stars.

Furthermore, the simulations of \citet{Myers14} form somewhat smaller systems (``clumps'') than the full cluster simulations of \citet{Dale12a,Dale13,Dale14}, and their feedback mechanisms also differ (for example, they include protostellar outflows but no photoionisation feedback or stellar winds). It is possible that this feedback is weaker in terms of regulating accretion flows, and hence the calculations of \citet{Myers14} may behave more like the control runs from \citet{Dale12a,Dale13,Dale14}, unless magnetic fields are dominant in governing the spatial distribution of stars. In either case, more simulations that include magnetic fields are required to address this question.

This suggests that either the advent of mass segregation in simulations of star formation is a random event, or that the prescriptions of feedback for the initial conditions differ enough to produce significantly different spatial distributions of stars.

In the SPH simulations of \citet{Dale12a,Dale13,Dale14} the runs without feedback lead to the most massive stars attaining much higher local densities than those with feedback. The reason for this is probably that feedback from the O-stars essentially shuts down accretion onto the main subclusters so they stop growing, both in the sense of the stars they already have not acquiring more mass, and in the sense of them having nothing to make new stars with. Therefore, the local SFE and thus local stellar density do not increase as much in the feedback runs as they do in the control runs. 

The initial conditions of the SPH runs (in terms of initial cloud mass, density and virial ratio) do not appear to influence whether mass segregation occurs, and certainly not as  much as the presence (or absence) of feedback. If we consider the $\mathcal{Q} - \Lambda_{\rm MSR}$ plot (Fig.~\ref{lambda_Q}), in the case without feedback (panel a) Run I is mass segregated, but the more dense version of this simulation (Run J) is not. Conversely, in the presence of feedback (panel b) Run I is not mass segregated, whereas Run J is.

Whilst resolution tests were conducted on some of the SPH simulations presented in \citet{Dale12a}, they were mainly directed at showing that the convergence of global properties such as the evolution of the star formation efficiency and global ionisation fraction was adequate, and did not include any of the simulations discussed here. The principal effect of increasing the resolution in an SPH simulation is to permit the formation of lower--mass objects which otherwise cannot be modelled. We did not find that the \emph{shape} of the mass function (except at the lowest--mass end), the total stellar mass, or the spatial distribution of stellar mass were substantially affected by the simulation resolution.

We also found no evidence that feedback alters the stellar mass function, so that increasing the simulation resolution would be likely to affect the stellar content of the control and feedback hydrodynamic simulations in very similar ways, by effectively extending their mass functions to lower masses. In reality some of this additional fragmentation would be prevented by physics which is not currently included in the SPH simulations, such as accretion feedback.

The effect of the presence of an additional population of low--mass stars on the outcome of the $N$-body simulations is not trivial to quantify. It is very unlikely to qualitatively alter our results (one of which is that it is not always possible to get a consensus from different algorithms on whether mass segregation is present in a single dataset), although it may alter the timescales on which the modelled stellar systems dynamically mass segregate or unsegregate. The timescale for mass segregation is related to the local crossing time (i.e.\,\,local density) and the presence of more low-mass stars would make the region more dense, reduce the crossing time and hence make the region more likely to dynamically mass segregate on faster timescales. However, very dense clusters can eject massive stars \citep[sometimes dissolving the cluster entirely,][]{Allison11,Parker14b}, so it is not clear whether the `missing' low-mass stars in the SPH runs would affect the amount of mass segregation measured.

During the subsequent dynamical evolution via $N$-body simulations, mass segregation can be erased through dynamical interactions or (more usually) mass-loss from stellar evolution, which has the effect of changing the ten most massive stars in the bin used to define $\Sigma_{\rm LDR}$ or $\Lambda_{\rm MSR}$.

Taken together, the above results suggest that mass segregation need not be primordial, and conversely, that a high-mass cluster which does not display mass segregation after a certain time may have been mass segregated in the past, especially if it contains highly evolved stars. The only limits we can place on mass segregation is that if it is not observed in a low-mass region without evolved stars \citep{Parker11b,Parker12c}, then it is unlikely that this region was ever mass segregated. Models that explain mass segregation of more massive clusters (such as the ONC) via the cool collapse of a star-forming region \citep{Allison10,Allison11,Parker14b} are still as likely an explanation for the observed mass segregation in these clusters, as opposed to primordial mass segregation.

\section{Conclusions}
\label{conclude}

We have evolved five triplets of hydrodynamical simulations of star formation from \citet{Dale12a,Dale13,Dale14} further in time using the $N$-body method. The hydrodynamical simulations follow the formation of a region with (a) no feedback, (b) photoionisation feedback only, and (c) photoionisation feedback and stellar winds. 

We have looked for differences in the spatial distribution of stars, local density and the fraction of bound stars. We have also used three different methods used to quantify mass segregation -- $\Lambda_{\rm MSR}$ \citep{Allison09a}, $\Sigma_{\rm LDR}$ \citep{Maschberger11,Parker14b} and the group segregation ratio, $\Omega_{\rm GSR}$ \citep{Kirk10,Kirk14}. 

In terms of the global evolution of the regions, our conclusions are similar to those in \citet{Parker13a}, where we looked for differences in simulations with no feedback, and simulations with photoionisation feedback only. The addition of stellar winds feedback has little effect above the photoionisation feedback; the star-forming regions subject to feedback form with lower densities and hence remain substructured for longer than the simulation without feedback. Generally, the simulations that form with feedback contain fewer bound stars at the end of the $N$-body integration, possibly due to the lower mass of the star-forming regions. 

Using three different techniques to search for mass segregation presents a rather confusing picture. When feedback is not switched on, the regions are usually primordially mass segregated according to $\Omega_{\rm GSR}$ and $\Sigma_{\rm LDR}$, whereas the $\Lambda_{\rm MSR}$ measure does not usually measure primordial mass segregation in the simulations. When feedback is switched on, all three measures show no preferential spatial distribution for the most massive stars, suggesting that the inclusion of more realistic physics in the hydrodynamical simulations suppresses primordial mass segregation, although more simulations that include magnetic fields would be highly desirable, and could potentially alter these conclusions.

When evolved for 10\,Myr, some simulations dynamically mass segregate. However, a combination of further dynamical evolution, and stellar evolution of the most massive stars, can also cause some regions to un-segregate, and sometimes re-segregate due to different subsets of massive stars being included in the determination. 

We conclude that extreme caution should be exercised when interpreting mass segregation in the outcome of hydrodynamical simulations of star formation. Different methods define mass segregation in different ways, and dynamical and stellar evolution can also affect searches of mass segregation. We note that several observed low-mass regions do not display mass segregation, and based on the results presented here, we suggest that mass segregation is not always primordial.

\section*{Acknowledgements}

RJP acknowledges support from the Royal Astronomical Society in the form of a research fellowship. This research was supported by the DFG cluster of excellence `Origin and Structure of the Universe' (JED, BE). We thank Tobias Aschenbrenner for the preliminary study in his Bachelorarbeit at USM, which was the genesis of this paper.

\bibliography{general_ref}

\begin{thebibliography}{}

\bibitem[\protect\citeauthoryear{Allison \& Goodwin}{Allison \&
  Goodwin}{2011}]{Allison11}
Allison R.~J.,  Goodwin S.~P.,  2011, MNRAS, 415, 1967

\bibitem[\protect\citeauthoryear{Allison, Goodwin, Parker, de Grijs, {Portegies
  Zwart} \& Kouwenhoven}{Allison et~al.}{2009}]{Allison09b}
Allison R.~J.,  Goodwin S.~P.,  Parker R.~J.,  de Grijs R.,  {Portegies Zwart}
  S.~F.,    Kouwenhoven M. B.~N.,  2009, ApJ, 700, L99

\bibitem[\protect\citeauthoryear{Allison, Goodwin, Parker, {Portegies Zwart} \&
  de Grijs}{Allison et~al.}{2010}]{Allison10}
Allison R.~J.,  Goodwin S.~P.,  Parker R.~J.,  {Portegies Zwart} S.~F.,    de
  Grijs R.,  2010, MNRAS, 407, 1098

\bibitem[\protect\citeauthoryear{Allison, Goodwin, Parker, {Portegies Zwart},
  de Grijs \& Kouwenhoven}{Allison et~al.}{2009}]{Allison09a}
Allison R.~J.,  Goodwin S.~P.,  Parker R.~J.,  {Portegies Zwart} S.~F.,  de
  Grijs R.,    Kouwenhoven M. B.~N.,  2009, MNRAS, 395, 1449

\bibitem[\protect\citeauthoryear{{Arzoumanian}, {Andr{\'e}}, {Didelon},
  {K{\"o}nyves}, {Schneider}, {Men'shchikov}, {Sousbie}, {Zavagno}, {Bontemps}
  \& {et al.}}{{Arzoumanian} et~al.}{2011}]{Arzoumanian11}
{Arzoumanian} D.,  {Andr{\'e}} P.,  {Didelon} P.,  {K{\"o}nyves} V.,
  {Schneider} N.,  {Men'shchikov} A.,  {Sousbie} T.,  {Zavagno} A.,  {Bontemps}
  S.,    {et al.} 2011, A\&A, 529, L6

\bibitem[\protect\citeauthoryear{{Ascenso}, {Alves} \& {Lago}}{{Ascenso}
  et~al.}{2009}]{Ascenso09}
{Ascenso} J.,  {Alves} J.,    {Lago} M.~T.~V.~T.,  2009, A\&A, 495, 147

\bibitem[\protect\citeauthoryear{Bastian, Gieles, Ercolano \&
  Gutermuth}{Bastian et~al.}{2009}]{Bastian09}
Bastian N.,  Gieles M.,  Ercolano B.,    Gutermuth R.,  2009, MNRAS, 392, 868

\bibitem[\protect\citeauthoryear{Bate}{Bate}{2009}]{Bate09}
Bate M.~R.,  2009, MNRAS, 392, 590

\bibitem[\protect\citeauthoryear{Bate}{Bate}{2012}]{Bate12}
Bate M.~R.,  2012, MNRAS, 419, 3115

\bibitem[\protect\citeauthoryear{{Baumgardt}, {Hut} \& {Heggie}}{{Baumgardt}
  et~al.}{2002}]{Baumgardt02}
{Baumgardt} H.,  {Hut} P.,    {Heggie} D.~C.,  2002, MNRAS, 336, 1069

\bibitem[\protect\citeauthoryear{Bonnell, Bate, Clarke \& Pringle}{Bonnell
  et~al.}{1997}]{Bonnell97}
Bonnell I.~A.,  Bate M.~R.,  Clarke C.~J.,    Pringle J.~E.,  1997, MNRAS, 285,
  201

\bibitem[\protect\citeauthoryear{{Bonnell}, {Bate} \& {Zinnecker}}{{Bonnell}
  et~al.}{1998}]{Bonnell98b}
{Bonnell} I.~A.,  {Bate} M.~R.,    {Zinnecker} H.,  1998, MNRAS, 298, 93

\bibitem[\protect\citeauthoryear{Bonnell, Clark \& Bate}{Bonnell
  et~al.}{2008}]{Bonnell08}
Bonnell I.~A.,  Clark P.~C.,    Bate M.~R.,  2008, MNRAS, 389, 1556

\bibitem[\protect\citeauthoryear{Bonnell \& Davies}{Bonnell \&
  Davies}{1998}]{Bonnell98}
Bonnell I.~A.,  Davies M.~B.,  1998, MNRAS, 295, 691

\bibitem[\protect\citeauthoryear{{Carpenter}, {Meyer}, {Dougados}, {Strom} \&
  {Hillenbrand}}{{Carpenter} et~al.}{1997}]{Carpenter97}
{Carpenter} J.~M.,  {Meyer} M.~R.,  {Dougados} C.,  {Strom} S.~E.,
  {Hillenbrand} L.~A.,  1997, AJ, 114, 198

\bibitem[\protect\citeauthoryear{Cartwright}{Cartwright}{2009}]{Cartwright09b}
Cartwright A.,  2009, MNRAS, 400, 1427

\bibitem[\protect\citeauthoryear{Cartwright \& Whitworth}{Cartwright \&
  Whitworth}{2004}]{Cartwright04}
Cartwright A.,  Whitworth A.~P.,  2004, MNRAS, 348, 589

\bibitem[\protect\citeauthoryear{{Cartwright} \& {Whitworth}}{{Cartwright} \&
  {Whitworth}}{2009}]{Cartwright09a}
{Cartwright} A.,  {Whitworth} A.~P.,  2009, MNRAS, 392, 341

\bibitem[\protect\citeauthoryear{Casertano \& Hut}{Casertano \&
  Hut}{1985}]{Casertano85}
Casertano S.,  Hut P.,  1985, ApJ, 298, 80

\bibitem[\protect\citeauthoryear{{Dale} \& {Bonnell}}{{Dale} \&
  {Bonnell}}{2008}]{Dale08}
{Dale} J.~E.,  {Bonnell} I.~A.,  2008, MNRAS, 391, 2

\bibitem[\protect\citeauthoryear{{Dale}, {Ercolano} \& {Bonnell}}{{Dale}
  et~al.}{2012}]{Dale12a}
{Dale} J.~E.,  {Ercolano} B.,    {Bonnell} I.~A.,  2012, MNRAS, 424, 377

\bibitem[\protect\citeauthoryear{{Dale}, {Ercolano} \& {Bonnell}}{{Dale}
  et~al.}{2013}]{Dale13}
{Dale} J.~E.,  {Ercolano} B.,    {Bonnell} I.~A.,  2013, MNRAS, 430, 234

\bibitem[\protect\citeauthoryear{{Dale}, {Ercolano} \& {Clarke}}{{Dale}
  et~al.}{2007}]{Dale07}
{Dale} J.~E.,  {Ercolano} B.,    {Clarke} C.~J.,  2007, MNRAS, 382, 1759

\bibitem[\protect\citeauthoryear{{Dale}, {Ngoumou}, {Ercolano} \&
  {Bonnell}}{{Dale} et~al.}{2014}]{Dale14}
{Dale} J.~E.,  {Ngoumou} J.,  {Ercolano} B.,    {Bonnell} I.~A.,  2014, MNRAS,
  442, 694

\bibitem[\protect\citeauthoryear{{de Grijs}, {Johnson}, {Gilmore} \&
  {Frayn}}{{de Grijs} et~al.}{2002}]{deGrijs02}
{de Grijs} R.,  {Johnson} R.~A.,  {Gilmore} G.~F.,    {Frayn} C.~M.,  2002,
  MNRAS, 331, 228

\bibitem[\protect\citeauthoryear{{Delgado-Donate}, Clarke, Bate \&
  Hodgkin}{{Delgado-Donate} et~al.}{2004}]{Donate04b}
{Delgado-Donate} E.~J.,  Clarke C.~J.,  Bate M.~R.,    Hodgkin S.~T.,  2004,
  MNRAS, 351, 617

\bibitem[\protect\citeauthoryear{{Eggleton}, {Fitchett} \& {Tout}}{{Eggleton}
  et~al.}{1989}]{Eggleton89}
{Eggleton} P.~P.,  {Fitchett} M.~J.,    {Tout} C.~A.,  1989, ApJ, 347, 998

\bibitem[\protect\citeauthoryear{{Girichidis}, {Federrath}, {Allison},
  {Banerjee} \& {Klessen}}{{Girichidis} et~al.}{2012}]{Girichidis12}
{Girichidis} P.,  {Federrath} C.,  {Allison} R.,  {Banerjee} R.,    {Klessen}
  R.~S.,  2012, MNRAS, 420, 3264

\bibitem[\protect\citeauthoryear{{Girichidis}, {Federrath}, {Banerjee} \&
  {Klessen}}{{Girichidis} et~al.}{2011}]{Girichidis11}
{Girichidis} P.,  {Federrath} C.,  {Banerjee} R.,    {Klessen} R.~S.,  2011,
  MNRAS, 413, 2741

\bibitem[\protect\citeauthoryear{Goodwin \& Bastian}{Goodwin \&
  Bastian}{2006}]{Goodwin06}
Goodwin S.~P.,  Bastian N.,  2006, MNRAS, 373, 752

\bibitem[\protect\citeauthoryear{Goodwin, Whitworth \& {Ward-Thompson}}{Goodwin
  et~al.}{2004}]{Goodwin04b}
Goodwin S.~P.,  Whitworth A.~P.,    {Ward-Thompson} D.,  2004, A\&A, 414, 633

\bibitem[\protect\citeauthoryear{Gouliermis, Keller, Kontizas, Kontizas \&
  {Bellas-Velidis}}{Gouliermis et~al.}{2004}]{Gouliermis04}
Gouliermis D.,  Keller S.~C.,  Kontizas M.,  Kontizas E.,    {Bellas-Velidis}
  I.,  2004, A\&A, 416, 137

\bibitem[\protect\citeauthoryear{{Gouliermis}, {de Grijs} \&
  {Xin}}{{Gouliermis} et~al.}{2009}]{Gouliermis09}
{Gouliermis} D.~A.,  {de Grijs} R.,    {Xin} Y.,  2009, ApJ, 692, 1678

\bibitem[\protect\citeauthoryear{Gutermuth, Megeath, Myers, Allen \&
  Fazio}{Gutermuth et~al.}{2009}]{Gutermuth09}
Gutermuth R.~A.,  Megeath S.~T.,  Myers P.~C.,  Allen L.~E.,    Fazio J. L. P.
  G.~G.,  2009, ApJS, 184, 18

\bibitem[\protect\citeauthoryear{{Haisch} Jr., {Lada} \& {Lada}}{{Haisch}
  et~al.}{2001}]{Haisch01}
{Haisch} Jr. K.~E.,  {Lada} E.~A.,    {Lada} C.~J.,  2001, ApJL, 553, L153

\bibitem[\protect\citeauthoryear{{Hansen}, {Klein}, {McKee} \&
  {Fisher}}{{Hansen} et~al.}{2012}]{Hansen12}
{Hansen} C.~E.,  {Klein} R.~I.,  {McKee} C.~F.,    {Fisher} R.~T.,  2012, ApJ,
  747, 22

\bibitem[\protect\citeauthoryear{{Hillenbrand}}{{Hillenbrand}}{1997}]{Hillenbr%
and97}
{Hillenbrand} L.~A.,  1997, AJ, 113, 1733

\bibitem[\protect\citeauthoryear{Hillenbrand \& Hartmann}{Hillenbrand \&
  Hartmann}{1998}]{Hillenbrand98}
Hillenbrand L.~A.,  Hartmann L.~W.,  1998, ApJ, 492, 540

\bibitem[\protect\citeauthoryear{Kirk \& Myers}{Kirk \& Myers}{2011}]{Kirk10}
Kirk H.,  Myers P.~C.,  2011, ApJ, 727, 64

\bibitem[\protect\citeauthoryear{{Kirk}, {Offner} \& {Redmond}}{{Kirk}
  et~al.}{2014}]{Kirk14}
{Kirk} H.,  {Offner} S.~S.~R.,    {Redmond} K.~J.,  2014, MNRAS, 439, 1765

\bibitem[\protect\citeauthoryear{Kruijssen}{Kruijssen}{2012}]{Kruijssen12b}
Kruijssen J. M.~D.,  2012, MNRAS, 426, 3008

\bibitem[\protect\citeauthoryear{Kruijssen, Maschberger, Moeckel, Clarke,
  Bastian \& Bonnell}{Kruijssen et~al.}{2012}]{Kruijssen12a}
Kruijssen J. M.~D.,  Maschberger T.,  Moeckel N.,  Clarke C.~J.,  Bastian N.,
   Bonnell I.~A.,  2012, MNRAS, 419, 841

\bibitem[\protect\citeauthoryear{Krumholz, McKee \& Klein}{Krumholz
  et~al.}{2005}]{Krumholz05}
Krumholz M.,  McKee C.~F.,    Klein R.,  2005, Nature, 438, 332

\bibitem[\protect\citeauthoryear{{Krumholz}, {Klein} \& {McKee}}{{Krumholz}
  et~al.}{2012}]{Krumholz12}
{Krumholz} M.~R.,  {Klein} R.~I.,    {McKee} C.~F.,  2012, \apj, 754, 71

\bibitem[\protect\citeauthoryear{{K{\"u}pper}, {Maschberger}, {Kroupa} \&
  {Baumgardt}}{{K{\"u}pper} et~al.}{2011}]{Kupper11}
{K{\"u}pper} A.~H.~W.,  {Maschberger} T.,  {Kroupa} P.,    {Baumgardt} H.,
  2011, MNRAS, 417, 2300

\bibitem[\protect\citeauthoryear{Lada \& Lada}{Lada \& Lada}{2003}]{Lada03}
Lada C.~J.,  Lada E.~A.,  2003, ARA\&A, 41, 57

\bibitem[\protect\citeauthoryear{Lada, Margulis \& Dearborn}{Lada
  et~al.}{1984}]{Lada84}
Lada C.~J.,  Margulis M.,    Dearborn D.,  1984, ApJ, 285, 141

\bibitem[\protect\citeauthoryear{McKee \& Tan}{McKee \& Tan}{2003}]{McKee03}
McKee C.~F.,  Tan J.~C.,  2003, ApJ, 585, 850

\bibitem[\protect\citeauthoryear{Maschberger \& Clarke}{Maschberger \&
  Clarke}{2011}]{Maschberger11}
Maschberger T.,  Clarke C.~J.,  2011, MNRAS, 416, 541

\bibitem[\protect\citeauthoryear{Moeckel \& Bate}{Moeckel \&
  Bate}{2010}]{Moeckel10}
Moeckel N.,  Bate M.~R.,  2010, MNRAS, 404, 721

\bibitem[\protect\citeauthoryear{{Moeckel} \& {Bonnell}}{{Moeckel} \&
  {Bonnell}}{2009}]{Moeckel09b}
{Moeckel} N.,  {Bonnell} I.~A.,  2009, MNRAS, 400, 657

\bibitem[\protect\citeauthoryear{Moeckel, Holland, Clarke \& Bonnell}{Moeckel
  et~al.}{2012}]{Moeckel12}
Moeckel N.,  Holland C.,  Clarke C.~J.,    Bonnell I.~A.,  2012, MNRAS, 425,
  450

\bibitem[\protect\citeauthoryear{{Myers}, {Klein}, {Krumholz} \&
  {McKee}}{{Myers} et~al.}{2014}]{Myers14}
{Myers} A.~T.,  {Klein} R.~I.,  {Krumholz} M.~R.,    {McKee} C.~F.,  2014,
  MNRAS, 439, 3420

\bibitem[\protect\citeauthoryear{Offner, Hansen \& Krumholz}{Offner
  et~al.}{2009}]{Offner09}
Offner S. S.~R.,  Hansen C.~E.,    Krumholz M.~R.,  2009, ApJ, 704, L124

\bibitem[\protect\citeauthoryear{{Offner}, {Kratter}, {Matzner}, {Krumholz} \&
  {Klein}}{{Offner} et~al.}{2010}]{Offner10}
{Offner} S.~S.~R.,  {Kratter} K.~M.,  {Matzner} C.~D.,  {Krumholz} M.~R.,
  {Klein} R.~I.,  2010, ApJ, 725, 1485

\bibitem[\protect\citeauthoryear{Parker, Bouvier, Goodwin, Moraux, Allison,
  Guieu \& G{\"u}del}{Parker et~al.}{2011}]{Parker11b}
Parker R.~J.,  Bouvier J.,  Goodwin S.~P.,  Moraux E.,  Allison R.~J.,  Guieu
  S.,    G{\"u}del M.,  2011, MNRAS, 412, 2489

\bibitem[\protect\citeauthoryear{Parker \& Dale}{Parker \&
  Dale}{2013}]{Parker13a}
Parker R.~J.,  Dale J.~E.,  2013, MNRAS, 432, 986

\bibitem[\protect\citeauthoryear{Parker, Maschberger \& {Alves de
  Oliveira}}{Parker et~al.}{2012}]{Parker12c}
Parker R.~J.,  Maschberger T.,    {Alves de Oliveira} C.,  2012, MNRAS, 426,
  3079

\bibitem[\protect\citeauthoryear{Parker, Wright, Goodwin \& Meyer}{Parker
  et~al.}{2014}]{Parker14b}
Parker R.~J.,  Wright N.~J.,  Goodwin S.~P.,    Meyer M.~R.,  2014, MNRAS, 438,
  620

\bibitem[\protect\citeauthoryear{{Peters}, {Klessen}, {Mac Low} \&
  {Banerjee}}{{Peters} et~al.}{2010}]{Peters10}
{Peters} T.,  {Klessen} R.~S.,  {Mac Low} M.-M.,    {Banerjee} R.,  2010, ApJ,
  725, 134

\bibitem[\protect\citeauthoryear{{Pinfield}, {Jameson} \& {Hodgkin}}{{Pinfield}
  et~al.}{1998}]{Pinfield98}
{Pinfield} D.~J.,  {Jameson} R.~F.,    {Hodgkin} S.~T.,  1998, MNRAS, 299, 955

\bibitem[\protect\citeauthoryear{{Portegies Zwart}, McMillan, Hut \&
  Makino}{{Portegies Zwart} et~al.}{2001}]{Zwart01}
{Portegies Zwart} S.~F.,  McMillan S. L.~W.,  Hut P.,    Makino J.,  2001,
  MNRAS, 321, 199

\bibitem[\protect\citeauthoryear{{Portegies Zwart}, Makino, McMillan \&
  Hut}{{Portegies Zwart} et~al.}{1999}]{Zwart99}
{Portegies Zwart} S.~F.,  Makino J.,  McMillan S. L.~W.,    Hut P.,  1999,
  A\&A, 348, 117

\bibitem[\protect\citeauthoryear{{Portegies Zwart} \& {Verbunt}}{{Portegies
  Zwart} \& {Verbunt}}{1996}]{Zwart96}
{Portegies Zwart} S.~F.,  {Verbunt} F.,  1996, A\&A, 309, 179

\bibitem[\protect\citeauthoryear{{Portegies Zwart} \& {Verbunt}}{{Portegies
  Zwart} \& {Verbunt}}{2012}]{Zwart12}
{Portegies Zwart} S.~F.,  {Verbunt} F.,  2012, Astrophysics Source Code
  Library, p.~1003

\bibitem[\protect\citeauthoryear{{Schmeja} \& {Klessen}}{{Schmeja} \&
  {Klessen}}{2006}]{Schmeja06}
{Schmeja} S.,  {Klessen} R.~S.,  2006, A\&A, 449, 151

\bibitem[\protect\citeauthoryear{Smith, Fellhauer, Goodwin \& Assmann}{Smith
  et~al.}{2011}]{Smith11}
Smith R.,  Fellhauer M.,  Goodwin S.,    Assmann P.,  2011, MNRAS, 414, 3036

\bibitem[\protect\citeauthoryear{{Spitzer} Jr.}{{Spitzer}}{1969}]{Spitzer69}
{Spitzer} Jr. L.,  1969, ApJL, 158, L139

\bibitem[\protect\citeauthoryear{{Tout}, {Pols}, {Eggleton} \& {Han}}{{Tout}
  et~al.}{1996}]{Tout96}
{Tout} C.~A.,  {Pols} O.~R.,  {Eggleton} P.~P.,    {Han} Z.,  1996, \mnras,
  281, 257

\bibitem[\protect\citeauthoryear{Tutukov}{Tutukov}{1978}]{Tutukov78}
Tutukov A.~V.,  1978, A\&A, 70, 57

\bibitem[\protect\citeauthoryear{Wright, Parker, Goodwin \& Drake}{Wright
  et~al.}{2014}]{Wright14}
Wright N.~J.,  Parker R.~J.,  Goodwin S.~P.,    Drake J.~J.,  2014, MNRAS, 438,
  639

\bibitem[\protect\citeauthoryear{{Zinnecker}}{{Zinnecker}}{1982}]{Zinnecker82}
{Zinnecker} H.,  1982, Annals of the New York Academy of Sciences, 395, 226

\end{thebibliography}

\label{lastpage}

\end{document}